\documentclass[a4paper, amsfonts, amssymb, amsmath, reprint, showkeys, nofootinbib, twoside, superscriptaddress]{revtex4-1}
\usepackage[english]{babel}
\usepackage[utf8]{inputenc}
\usepackage{float}
\usepackage{amsthm}
\usepackage{mathtools}
\usepackage{physics}
\usepackage{xcolor}
\usepackage{graphicx}
\usepackage[left=23mm,right=13mm,top=35mm,columnsep=15pt]{geometry} 
\usepackage{adjustbox}
\usepackage{placeins}
\usepackage[T1]{fontenc}
\usepackage{lipsum}
\usepackage{csquotes}
\usepackage[pdftex, pdftitle={Article}, pdfauthor={Author}]{hyperref} 

\begin{document}

\title{\textbf{Experimental probe of a complete 3D photonic band gap}}  

\author{Manashee Adhikary}
\affiliation{Complex Photonic Systems (COPS), 
MESA+ Institute for Nanotechnology, 
University of Twente, 
P.O. Box 217, 7500 AE Enschede, The Netherlands.}
\author{Ravitej Uppu}
\affiliation{Complex Photonic Systems (COPS), 
MESA+ Institute for Nanotechnology, 
University of Twente, 
P.O. Box 217, 7500 AE Enschede, The Netherlands.}
\affiliation{Present address: Center for Hybrid Quantum Networks (Hy-Q), 
Niels Bohr Institute, University of Copenhagen, 
Blegdamsvej 17, 2100-DK Copenhagen, Denmark }

\author{Cornelis A.M. Harteveld}
\affiliation{Complex Photonic Systems (COPS), 
MESA+ Institute for Nanotechnology, 
University of Twente, 
P.O. Box 217, 7500 AE Enschede, The Netherlands.}

\author{Diana A. Grishina}
\affiliation{Complex Photonic Systems (COPS), 
MESA+ Institute for Nanotechnology, 
University of Twente, 
P.O. Box 217, 7500 AE Enschede, The Netherlands.}
\affiliation{Present address: ASML Netherlands B.V. (HQ), 
De Run 6501, 5504 DR Veldhoven, The Netherlands}

\author{Willem L. Vos}
\affiliation{Complex Photonic Systems (COPS), 
MESA+ Institute for Nanotechnology, 
University of Twente, 
P.O. Box 217, 7500 AE Enschede, The Netherlands.}
\email{w.l.vos@utwente.nl} 

\homepage{www.photonicbandgaps.com} 

\begin{abstract}
The identification of a complete three-dimensional (3D) photonic band gap in real crystals always employs theoretical or numerical models that invoke idealized crystal structures. 
Thus, this approach is prone to false positives (gap wrongly assigned) or false negatives (gap missed). 
Therefore, we propose a purely experimental probe of the 3D photonic band gap that pertains to many different classes of photonic materials. 
We study position and polarization-resolved reflectivity spectra of 3D inverse woodpile structures that consist of two perpendicular nanopore arrays etched in silicon.
We observe intense reflectivity peaks $(R > 90\%)$ typical of high-quality crystals with broad stopbands. 
We track the stopband width versus pore radius, which agrees much better with the predicted 3D photonic band gap than with a directional stop gap on account of the large numerical aperture used. 
A parametric plot of s-polarized versus p-polarized stopband width agrees very well with the 3D band gap and is model-free. 
This practical probe provides fast feedback on the advanced nanofabrication needed for 3D photonic crystals and stimulates practical applications of band gaps in 3D silicon nanophotonics and photonic integrated circuits, photovoltaics, cavity QED, and quantum information processing.
\end{abstract}
\maketitle


\section{Introduction}
\begin{figure}[ht]
\centering
\includegraphics[width=70mm]{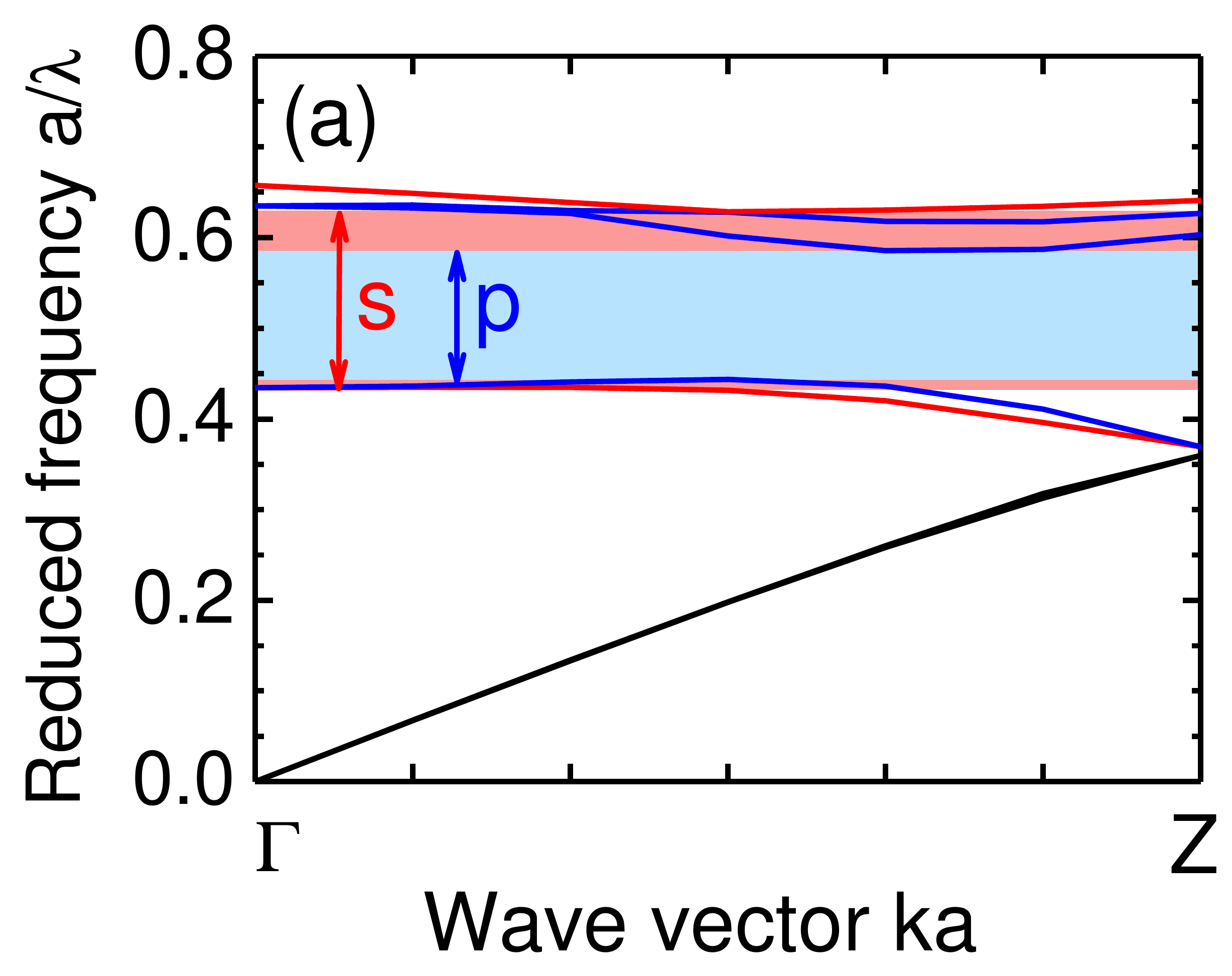} 
\includegraphics[width=70mm]{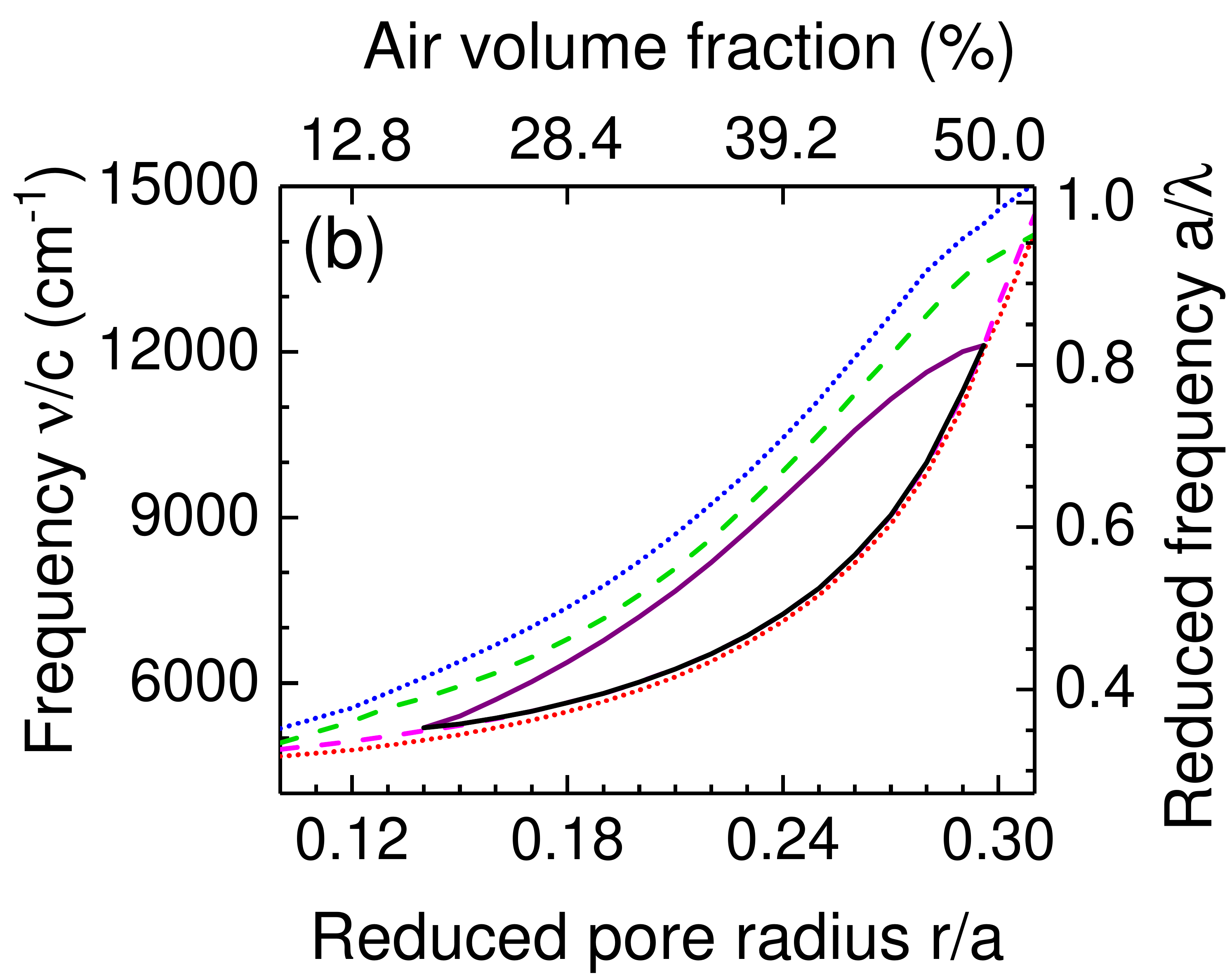} 
\caption{(a) Band structure of an inverse woodpile photonic crystal calculated for a reduced pore radius $r/a$ = 0.22 and a relative permittivity $\varepsilon_{Si}=11.68$. 
The abscissa is the reduced wave vector in the $\Gamma Z$ high-symmetry direction. 
The stop gaps for $s$-polarized and $p$-polarized light are indicated by the red and blue bars, respectively, and $p$-polarized bands are shown in blue and $s$ bands in red~\cite{Devashish2017PRB}. 
(b) The $\Gamma Z$ stop gap and photonic band gap as a function of the reduced pore radius $r/a$, with corresponding air volume fractions shown on the top abscissa. 
For $s$ and $p$ polarizations, the $\Gamma Z$ stop gap edges are shown as the blue and red dotted curves and the green and red dashed curves, respectively. 
The full black curves are the edges of the 3D band gap.} 
\label{Fig:bandstruct}
\end{figure}

Completely controlling the emission and the propagation of light simultaneously in all three dimensions (3D)  remains a major outstanding target in the field of Nanophotonics~\cite{Novotny2006book, Joannopoulos2008PhotonicLight, Lourtioz2008book, Noginov2009book, Ghulinyan2015book}. 
Particularly promising tools for this purpose are 3D photonic crystals with spatially periodic variations of the refractive index commensurate with optical wavelengths. 
The photon dispersion relations inside such crystals are organized in bands, analogous to electron bands in solids~\cite{Ashcroft1976book, Economou2010book}, see for example Figure~\ref{Fig:bandstruct}(a). 
When light waves inside a crystal are Bragg diffracted, directional energy gaps -- known as stop gaps -- arise for the relevant incident wavevector.  
When the stop gaps have a common overlap range for all wavevector and all polarizations, the 3D nanostructure has a photonic band gap. 
Within the band gap, no light modes are allowed in the crystal due to multiple Bragg interference~\cite{vanDriel2000PRB, Vos2000PLA, Romanov2001PRE}, hence the density of states (DOS) strictly vanishes. 
Since the local density of states also vanishes, the photonic band gap is a powerful tool to radically control spontaneous emission and cavity quantum electrodynamics (QED) of embedded quantum emitters~\cite{Bykov1972JETP, Yablonovitch1987PRL, John1990PRL, Vos2015CavityCrystals}. 
Applications of 3D photonic band gap crystals range from dielectric reflectors for antennae~\cite{Smith1998MOTL} and for efficient photovoltaic cells~\cite{Bermel2007OE, Wehrspohn2012JO, Koenderink2015science}, via white light-emitting diodes~\cite{David2012RPP}, to elaborate 3D waveguides~\cite{Li2003JOSA} for 3D photonic integrated circuits~\cite{Tajiri2019Optica}, and to thresholdless miniature lasers~\cite{Tandaechanurat2011NP} and devices to control quantum noise for quantum measurement, amplification, and information processing~\cite{Clerk2010RMP, Vos2015CavityCrystals}. 

Thanks to extensive research efforts in nanotechnology, great strides have been made in the fabrication of 3D nanostructures that interact strongly with light such that they possess a 3D full and complete photonic band gap~\cite{Lopez2003AM, Benisty2006ProgrOptics, Galisteo2011AM, Vos2015CavityCrystals}.
Remarkably, however, it remains a considerable challenge to decide firstly whether a 3D nanostructure has a \textit{bona fide} photonic band gap functionality or not, and secondly to assess how broad such a band gap is, which is critical for the robustness of the functionality.
It is natural to try to probe the photonic band gap via its influence on the DOS and LDOS by means of emission spectra or time-resolved emission dynamics of emitters embedded inside the photonic crystal~\cite{Ogawa2004S, Lodahl2004Nat, Aoki2008NP, Leistikow2011PRL}. 
However, such experiments are rather difficult for several practical reasons, that notably involve the emitter's quantum efficiency~\cite{Koenderink2002PRL}, the choice of a suitable reference system~\cite{Koenderink2003PSSA}, and finite-size effects~\cite{Hasan2018PRL}. 

Alternatively, the presence of a gap in the density of states may be probed by transmission or reflectivity~\cite{Lin1998Nature, Thijssen1999PRL, Noda2000Science, Blanco2000Nature, Vlasov2001N, Schilling2005APL, Garcia-Santamaria2007AM, Takahashi2009NM, Staude2010OL, Huisman2011PRB, Frolich2013AM, Marichy2016SR}. 
In such an experiment, a peak in reflectivity or a trough in transmission identifies a stopband in the real and finite crystal that is interpreted with a directional stop gap in the dispersion relations. 
By studying the 3D crystal over a sufficiently large solid angle, one expects to see a signature of a 3D photonic band gap. 
While reflectivity and transmission are readily measured, such probes suffer from two main limitations. 
One technical impediment is when a reflectivity or transmission experiment samples a too small angular range to safely assign a gap, whereas a broader range would reveal band overlap.
A second class of impediment includes possible artifacts related to uncoupled modes~\cite{Robertson1992PRL, Sakoda2005book}, fabrication imperfections, or unavoidable random disorder, all of which may lead either to erroneously assigned band gaps (`false positive') or to overlooked gaps (`false negative'). 
To date, these issues are addressed by supplementing reflectivity or transmission experiments with theoretical or numerical results and deciding the presence of a band gap and its width from such results. 
Theory or numerical simulations, however, always require a model for the photonic crystal's structure and the building blocks inside the unit cell.
Such a model is necessarily an idealization of the real crystal structure and thus misses essential features. 
For instance, crystal models are often taken to be infinitely extended and then lack an interface that essentially determines reflectivity features~\cite{Devashish2017PRB}. 
Or unavoidable disorder is not taken into account, while a certain degree of disorder may completely close a band gap~\cite{Li2000PRB}. 
Or the crystal structure model lacks random stacking (occurring in self-organized structures) which affects the presence and width of a band gap~\cite{Wang2003PRE}. 
Thus, in case that the ideal model differs from the real structure, the optical functionality of the crystal differs from the expected design for reasons that are far from trivial to identify~\cite{Grishina2018Arxiv}. 
Therefore, the goal of this paper is to find a purely experimental identification of a photonic band gap, which is robust to artifacts as it avoids the need for modeling. 
To this end, we collect polarization and position-resolved reflectivity spectra with a large numerical aperture. 
By mapping the width of the observed stopband versus a characteristic structural feature (here: pore radii in inverse woodpile crystals) that tunes the average refractive index, and by parametrically plotting the width of the observed s-polarized stopband versus the p-polarized one, we arrive at an experimental probe to decide whether a photonic crystal has a band gap.

\section{\label{sec:Methods}Samples and experimental}
\subsection{Inverse woodpile crystals}
Here we study 3D photonic band gap crystals with the inverse woodpile crystal structure~\cite{Ho1994SSC} made of silicon by CMOS-compatible means. 
The inverse woodpile structure is designed to consist of two identical two dimensional (2D) arrays of pores with radius $R$ running in the perpendicular $X$ and $Z$ directions.
Each 2D array of pores has a centered-rectangular structure with lattice constants $a$ and $c$ in a ratio $a/c=\sqrt{2}$ for the crystal structure to be cubic with a diamond-like symmetry, as illustrated in a YouTube animation~\cite{COPS2012youtube}. 
Inverse woodpile crystals have a broad 3D photonic band gap on account of their diamond-like structure~\cite{Maldovan2004NM} with a maximum relative bandwidth of 25.4\% for a reduced pore radius $r/a=0.245$ and a relative permittivity $\epsilon_{Si}=11.68$ typical of silicon backbone~\cite{Hillebrand2003JAP, Woldering2009JAP}. 

Figure~\ref{Fig:bandstruct}(a) shows the band structure calculated for the $\Gamma Z$ high symmetry direction since in our experiments the axis of the incident light cone is along this direction. 
The stop gap is the frequency range where modes are forbidden in this high symmetry direction. 
The relative bandwidth of the stop gap, gauged as the gap width $\Delta \omega$ to mid-gap $\omega_c$ ratio, is wider for $s$-polarized light ($\Delta \omega/\omega_c = 36.5\%$) than for $p$-polarized light ($\Delta \omega/\omega_c = 27.6\%$), which is reasonable since in the former case the electric field is perpendicular to the first layer of pores so that light scatters more strongly from this layer. 
For the diamond-like inverse woodpile structure, the $\Gamma Z$ high-symmetry direction is equivalent to the $\Gamma X$ high-symmetry direction, and thus also their opposite counterparts \textit{viz.} the $-\Gamma Z$ and $-\Gamma X$ high-symmetry directions~\cite{Huisman2011PRB, Devashish2017PRB}.
Several bands have $s$ or $p$-polarized character following the assignment of Ref.~\cite{Devashish2017PRB}. 
We refer to Bloch mode polarization to indicate their symmetry properties while being excited with either $s$ or $p$-polarized light incident from a high-symmetry direction (here the Z-direction). 

Figure~\ref{Fig:bandstruct}(b) shows the $\Gamma Z$ stop gaps for $s$ and $p$ polarization as a function of the reduced pore radius $r/a$, as well as the photonic band gap~\cite{Huisman2011PRB}. 
The centers of all gaps shift to higher frequencies which makes sense, 
since a gap center frequency $\omega_c$ is equal to $\omega_c = \frac{c'}{n_{eff}}.k_{BZ}.G$~\cite{Vos1996PRB}, with $c'$ the speed of light (not to be confused with the lattice parameter $c$), $n_{eff}$ the effective refractive index of the photonic crystal~\cite{Datta1993PRB}, and $G$ a structure-specific constant~\cite{Vos1996PRB}. 
An increasing pore radius in Fig.~\ref{Fig:bandstruct}(b) corresponds to an increasing air volume fraction and thus to a decreasing effective refractive index, hence to an increasing gap center frequency. 
As reported earlier, the 3D photonic band gap is the widest for $r/a=0.245$ and it is robust as it is open within the broad range $0.14 < r/a < 0.29$~\cite{Hillebrand2003JAP, Woldering2009JAP}.
When comparing the stop gaps and the 3D photonic band gap, we note that all lower edges nearly overlap, whereas the upper edges are all different.
The overlap of the lower edges of the stop gaps and the band gap is robust as a function of pore radius $(r/a)$ and hence effective refractive index, which is a convenient feature that we will exploit. 

\begin{figure}[htbp]
\centering
\includegraphics[width=66mm]{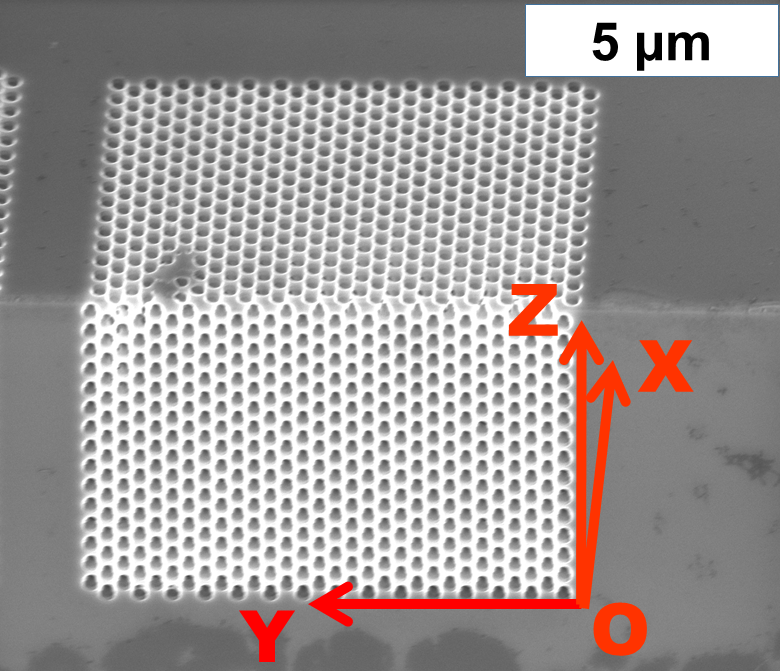}
\caption{Scanning electron microscopy (SEM) image of the edge of silicon beam \textbf{A} with a cubic 3D inverse woodpile crystal in perspective view.
The crystal consists of two sets of perpendicular pores along the $X$ and $Z$ directions with design radius $r_{d} = 160$ nm.
The coordinate system used in the paper is shown with the origin at the lower right corner of the crystal.
The crystal has lattice parameters $a = 680$ in the Y-direction, and $c$ in the X and Z-directions, with  $c=a/\sqrt{2}$~\cite{VanDenBroek2012AFM,Grishina20173DNanophotonics}.
}
\label{fig:SEM}
\end{figure}

The crystals are fabricated by etching pores into crystalline silicon using CMOS-compatible methods~\cite{VanDenBroek2012AFM}.
We employed deep reactive ion etching through an etch mask that was fabricated on the edge of a silicon beam~\cite{Tjerkstra2011JVSTB, Grishina2015Nanotech, Grishina20173DNanophotonics}. 
Multiple crystals with different design pore radii $r_{d}$ and a constant lattice parameter $a = 680$ nm were fabricated on a silicon beam.
One silicon beam, called \textbf{A}, contains eleven 3D crystals. 
We also present results obtained with another experimental setup on an older silicon beam \textbf{B} with five similar 3D crystals~\cite{Grishina20173DNanophotonics}.
Figure~\ref{fig:SEM} shows a scanning electron microscopy (SEM) image of one of our crystals with designed pore radius $r_d = 160$~nm ($r_d/a = 0.235$) on the edge of the silicon beam \textbf{A}.
The dimensions of each crystal are typically $8 \times 10 \times 8 \mu$m$^3$. 
Figure~\ref{fig:SEM} shows that the sample geometry allows for good optical access to the $XY$ and $YZ$ crystal surfaces.

\subsection{Near-infrared reflectivity microscope}

\begin{figure}[ht ]
\centering
\includegraphics[width=80mm]{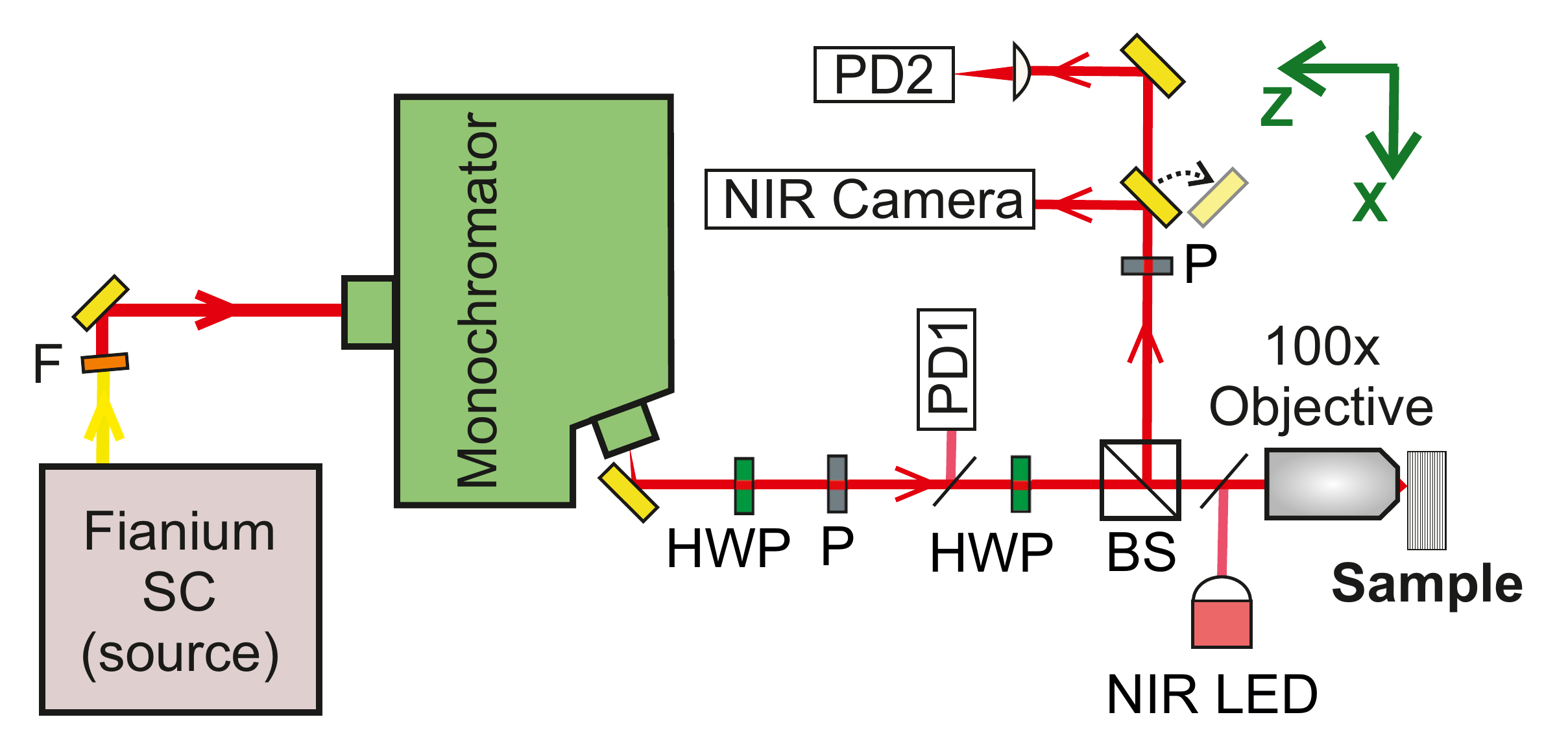}
\caption{Setup to measure position-resolved microscopic broadband reflectivity. 
The Fianium SC is the broadband supercontinuum source, the long-pass glass filter F blocks the visible light at $\lambda < 850$ nm, the monochromator filters the light to a narrow band, 
HWP are half-wave plates, P are polarizers, and BS are beam splitters. 
Incident light is focused on the sample with a 100$\times$ objective that also collects the reflected light; the coordinate system is shown at top right. 
The NIR camera views the sample in reflection with a magnification of 250$\times$. 
The photodiodes PD1 and PD2 monitor the incident light power and measure signal from the crystal, respectively.
}
\label{fig:setup}
\end{figure}

We have developed a near-infrared microscope setup to collect position-resolved broadband reflectivity spectra of photonic nanostructures, as is shown in Figure~\ref{fig:setup}. 
The near-infrared range of operation is compatible with 3D silicon nanophotonics as it allows to avoid the intrinsic silicon absorption.
The setup was developed with the option to collect in future light scattered perpendicular to the incident light.
Furthermore, a spatial light modulator can be inserted to eventually perform wavefront shaping~\cite{Vellekoop2007OptLt, Mosk2012NP}. 
Therefore, we decided to use sequential scanning of wavelengths instead of measuring the spectrum at once with a spectrometer as in Refs.~\cite{Ctistis2010PRB, Huisman2011PRB}.

In the optical setup shown in Figure~\ref{fig:setup}, the silicon beam with the 3D crystals is mounted on an XYZ translation stage that has a step size of about $30$ nm.
We use a broadband supercontinuum source (Fianium SC 400-4, 450 nm - 2400 nm) whose output is filtered by a long pass glass filter (Schott RG850) to block the unused visible range.
The near infrared light is spectrally selected by a monochromator (Oriel MS257; 1200 lines/mm grating) with an output linewidth of about $\Delta \lambda = 1$ nm and a tuning precision better than 0.2 nm. 
The accessible range of wavelengths spans from 900 nm to 2120 nm (or wave numbers $\nu/c = 11000$ cm$^{-1}$ to $4700$ cm$^{-1}$) in the near infrared including the telecom bands.
Using a combination of a linear polarizer and half wave plates, the linear polarization of the spectrally filtered light is selected and sent to an infrared apochromatic objective (Olympus MPlan Apo 100$\times$) to focus the light onto the sample's $XY$ surface with a numerical aperture NA $=0.85$.
The glass objective allows for access over the whole numerical aperture, instead of a blocked range around the axis as previously with a Schwarzschild reflecting objective~\cite{Ctistis2010PRB, Huisman2011PRB}.
The NA corresponds to a collection solid angle of $0.95\pi$ sr. 
On account of the crystal symmetry mentioned above ($\Gamma Z$ equivalent with $\Gamma X$ and with the opposite counterparts), we effectively collect a solid angle of $3.8\pi$ sr.

\begin{figure}[ht]
\centering
\includegraphics[width=60mm]{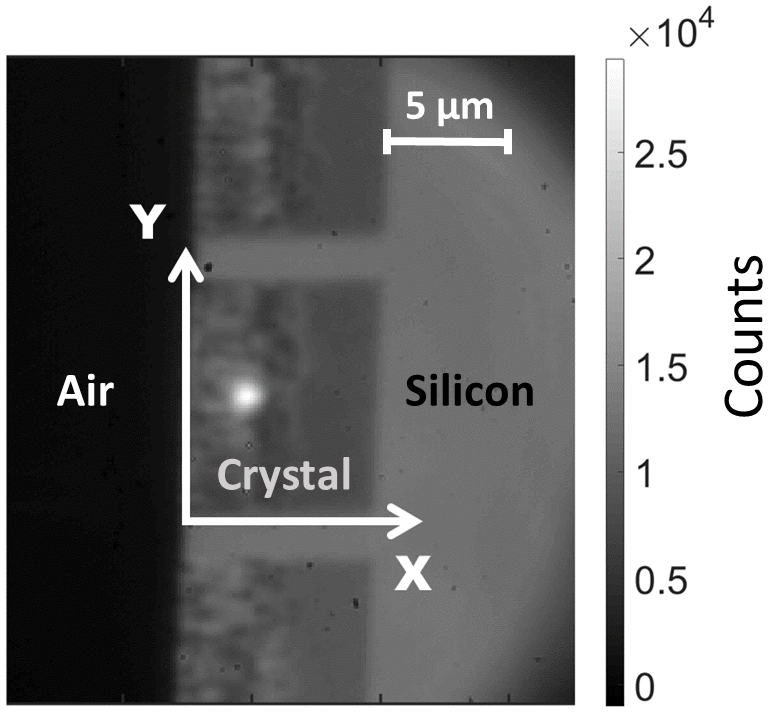}
\caption{Image of the $XY$-surface of one of the 3D inverse woodpile crystals on beam \textbf{A} as seen with the IR camera in the setup.
The bright spot is the focus of the incident light from the supercontinuum source filtered by the monochromator. 
The surface of the Si beam with the crystals is illuminated with a near infrared LED.
}
\label{fig:camera_image}
\end{figure}

Light reflected by the sample is collected by the same objective as shown in Figure \ref{fig:setup}.
A beam splitter directs the reflected light towards the detection arm where the reflection from the sample is imaged onto an IR camera (Photonic Science InGaAs).
In order to locate the focus of the input light on the surface, a near infrared LED is used to illuminate the sample surface.
We use the XYZ translation stage to move the sample to focus the light on the desired location.
An image as seen on the IR camera (see Fig.~\ref{fig:camera_image}) reveals the $XY$ surface of the Si beam.
The bright circular spot with a diameter of about 2 $\mu$m is the focus of light reflected from the crystal. 
The rectangular darker areas of about 8 $\mu$m $\times$10 $\mu$m are the XY surfaces of the 3D photonic crystals.
They appear dark compared to the surrounding silicon since the LED illumination is outside the band gap of these crystals whose effective refractive index is less than that of silicon.

Once the input light beam is focused on the sample, the reflected light is sent to photodiode PD2 (Thorlabs InGaAs DET10D/M, 900 nm - 2600 nm) by flipping off the mirror in front of the camera.
The photodiode records the reflected intensity $I_R$ as the monochromator scans the selected wavelength range.
An analyzer in front of the detector selects the polarization of the reflected light.
All reflectivity measurements are done for two orthogonal polarization states of the incident light, namely $s$ (electric field transverse to X-directed pores) and $p$ (electric field parallel to X-directed pores).
A typical spectrum takes about 5 to 25 minutes to record depending on the chosen wavelength step size of typically 10 nm or 2 nm.
Using the translation stage, the sample is moved in the Y-direction to select different crystals on the edge of the silicon beam. 

To calibrate the reflectivity defined as $R \equiv I_R/I_0$, the spectral response $I_R$ of the crystals is referenced to the signal $I_0$ from a clean gold mirror that reflects $96\%$. 
Calibration also removes dispersive contributions from optical components in the setup.
We ensure that the signal to noise ratio of the photodiode response is sufficient to detect signal in the desired range.
Therefore, the detector photodiode is fed into a lock-in amplifier to amplify the signal with a suitable gain. 
Since a serial measurement mode holds the risk of possible temporal variations in the supercontinuum source, we simultaneously collect the output of the monochromator with photodiode PD1 in each reflectivity scan. 
This monitor spectrum is used to normalize out variations in the incident intensity $I_0$. 
Since it is tedious to dismount and realign the sample to take reference spectra during a position scan, we also take secondary reference measurements on bulk silicon outside the crystals which has a flat response $R = 31\%$ with respect to the gold mirror.

We also discuss data measured on similar silicon beam \textbf{B} and obtained with an older setup employing a Fourier transform spectrometer and a Schwarzschild reflecting objective \cite{Grishina20173DNanophotonics, Huisman2011PRB, Ctistis2010PRB}. 
The maximum reflectivities are lower than in the new setup (30\% versus 90\%) probably on account of a larger spot size in this setup (compared to Refs.~\cite{Huisman2011PRB,Ctistis2010PRB} we find that the focus diameter has over the years changed from $1$ to $5 \mu$m.) 
Nevertheless, the measured peak positions and bandwidths agree well with the newer ones. 

\section{Results}
\subsection{\label{sec:stopbands}Stopbands}

\begin{figure}[ht]
\centering
\includegraphics[width=80mm]{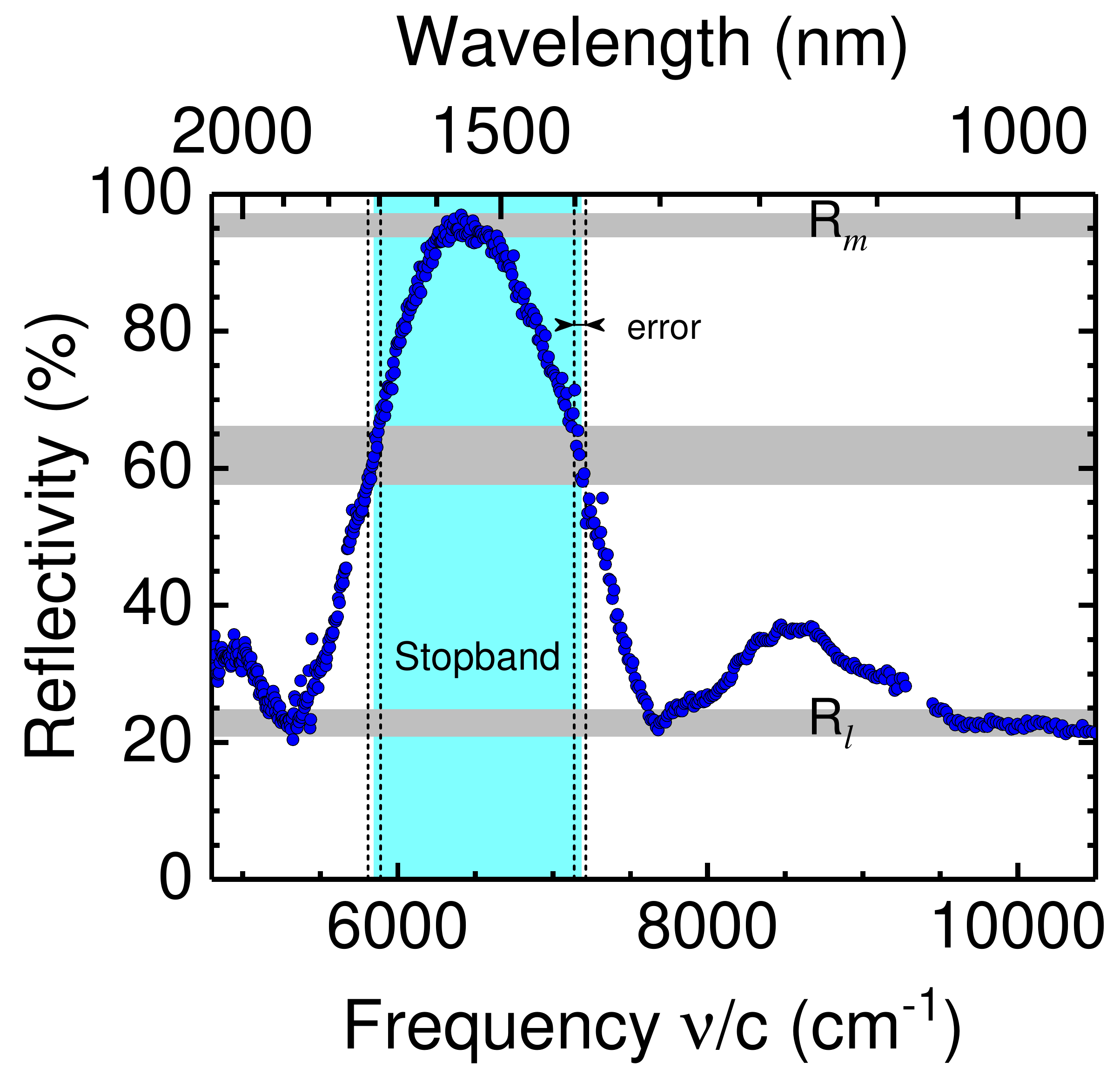}
\caption{Measured reflectivity spectrum of a 3D crystal with designed pore radius $r_d = 130$ nm ($r_d/a=0.191$) on silicon beam \textbf{A} for $p$-polarized input light. 
The stopband is estimated as the full-width at half-maximum of the reflectivity peak, shown by the cyan area.
The baseline reflectivity ($r_l$), maximum reflectivity ($r_m$) and half maximum are shown as the grey bars. 
The estimated error for both stopband edges is shown as the vertical dashed lines. 
}
\label{fig:determine_stopband}
\end{figure}

Figure~\ref{fig:determine_stopband} shows a reflectivity spectrum of a crystal with design pore radius $r_d = 130$ nm ($r_d/a=0.191$) recorded using our new setup. 
The broad and bright peak is the stopband that is associated with the main $\Gamma Z$ stop gap centered near $a/\lambda = 0.5$ in Figure~\ref{Fig:bandstruct}(a). 
The stopband width is taken as the full width at half maximum (FWHM) of the reflectivity peak~\cite{Vos2001NATO}.
The baseline of the peak is taken as the minimum reflectivity in the long-wavelength limit at frequencies below the stopband, with the standard deviation in this frequency range as the error margin. 
Similarly, the maximum reflectivity is taken as the mean in a narrow range around the peak, with the standard deviation in this range taken as the error margin. 
The baseline, the maximum reflectivity, and the half maximum are shown in Figure~\ref{fig:determine_stopband} as grey bars including their estimated errors.
The errors are propagated into the estimates of the edges at half maximum of the peak. 

\begin{figure}[ht]
\centering
\includegraphics[width=80mm]{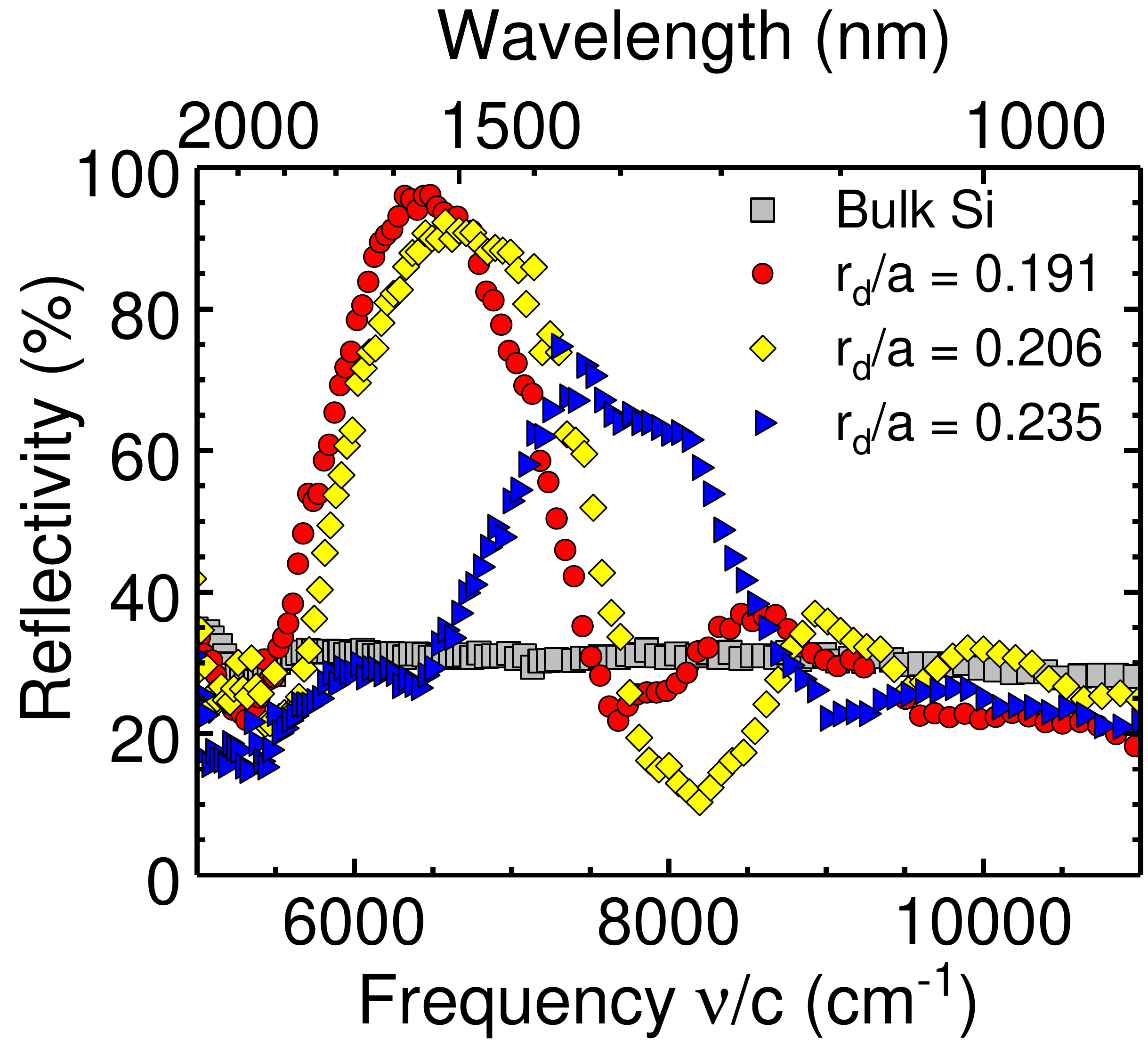}
\caption{Reflectivity spectra of three different 3D photonic crystals on Si beam \textbf{A} with three designed pore radii $r_d = 130, 140$ and $160$~nm $(r_d/a = 0.191, 0.206, 0.235)$ (red circles, yellow diamonds and blue triangles, respectively). 
The stopbands appear at different frequency ranges. 
The gray squares represent reflectivity from bulk Si on the beam away from the crystals. 
}
\label{fig:spectra_s9_s13_s18_si}
\end{figure}

Figure~\ref{fig:spectra_s9_s13_s18_si} shows reflectivity spectra measured on three 3D crystals on beam \textbf{A} with different designed pore radii $r_d = 130, 140, 160$~nm, as well as on the Si substrate.
Here, a change in the ratio of pore radius to the lattice constant $r_d/a$, called as the reduced pore radius corresponds to a change in the pore radius only since the lattice constants in our crystals are kept constant at 680 nm.
The constant reflectivity $R = 30.6 \pm {1.3} \%$ of the substrate agrees well with the Fresnel reflectivity of $31 \%$ expected for bulk silicon at normal incidence~\cite{NSM}. 
Intense reflectivity peaks with maxima of $R_{m} = 96\%$ and $94\%$ are measured on the crystals with pore radii $r_{d} = 130$~nm and $140$~nm, respectively. 
Our observations are consistent with recent numerical results that perfect silicon inverse woodpile crystals with a thickness of only three unit cells reflect $99 \%$ of the incident light~\cite{Devashish2017PRB}.
The results are also consistent with $95 \%$ reflectivity on a direct silicon woodpile that was only one unit cell thick by Euser \textit{et al.}~\cite{Euser2008PRB}.
We surmise that the current maximum reflectivities are higher than our previous results~\cite{Huisman2011PRB, Grishina2015Nanotech} due to improved nanofabrication and an improved optical setup. 
Figure \ref{fig:spectra_s9_s13_s18_si} also shows that the center of the stopband shifts to higher frequencies with increasing pore radius.
Such tuning of the stopband center with increasing pore radius qualitatively agrees with the behavior of the calculated stop gap and band gap shown in Figure~\ref{Fig:bandstruct}(b). 

The central question regarding reflectivity spectra as shown in Figure~\ref{fig:spectra_s9_s13_s18_si},  is which feature of a measured reflectivity peak is representative of characteristic photonic crystal features, such as a (directional) photonic stop gap or a (omnidirectional) photonic band gap. 
In case of weakly interacting photonic crystals, Ref.~\cite{Vos2001NATO} argued that the FWHM of a stopband collected with a low numerical aperture is a robust measure of a stop gap that is associated with one wave vector. 
Since such crystals weakly interact with light, there is a slim chance to find a photonic band gap. 
Using strongly interacting Si inverse opals, Palacios-Lid{\'{o}}n \textit{et al.} discussed that reflectivity collected over multiple high symmetry directions reveals a feature that is representative of the photonic band gap~\cite{PalaciosLidon2002APL}. 
Huisman \textit{et al.} proposed to combine measurements over several high-symmetry directions with a large numerical aperture since the band gap is associated with all wave vectors, hence $4 \pi$ sr solid angle~\cite{Huisman2011PRB}. 
Here, we propose to extend these earlier probes by mapping stopbands for $s$ and $p$-polarized light as a function of a structural parameter, \textit{viz.} the variation of the pore radii $r/a$, that entails the tuning of the effective refractive index. 

\subsection{Track pore radii from position-dependent stopband}\label{sec:position_dependent_stopbands}
To realize the mapping described above, we first identify a way to scan the pore radii. 
It is well-known from structural studies such as scanning electron microscopy on cleaved or milled crystals~\cite{VanDenBroek2012AFM} and from non-destructive traceless X-ray tomography~\cite{Grishina2018Arxiv}) that the radius of etched nanopores varies slightly around the designed value with position inside the crystal. 
By comparing the lower edge of the measured stopband with the calculated stop gap (\textit{cf.} Figure~\ref{Fig:bandstruct}(b)), we obtain an estimate of the local average pore radius $r$ at the position $\vec{\mathbf{r}}$ of the optical focus: $r (\vec{\mathbf{r}})$. 
In this comparison we take advantage of the feature in the band structures of inverse woodpile crystals that the lower edges of both the band gap and of the stop gap are nearly the same, see Figure~\ref{Fig:bandstruct}(b), hence the determination is robust to the interpretation which gap is probed. 
For the three crystals in Figure~\ref{fig:spectra_s9_s13_s18_si}, we derive the pore radii to be $r/a = 0.190\pm 0.001, 0.195\pm 0.001,~\textrm{and}~ 0.228\pm 0.002$, respectively, which agrees very well with the design ($r_d/a = 0.191,~0.206,~0.235$), where the small differencess are attributed to the depth-dependent pore radius discussed above.
We note that since the probing direction is perpendicular to the $X$-directed pores in the crystals, the derived pore radii are effectively those of the pores that run in the $X$-direction.

\begin{figure}[ht]
\centering
\includegraphics[width=80mm]{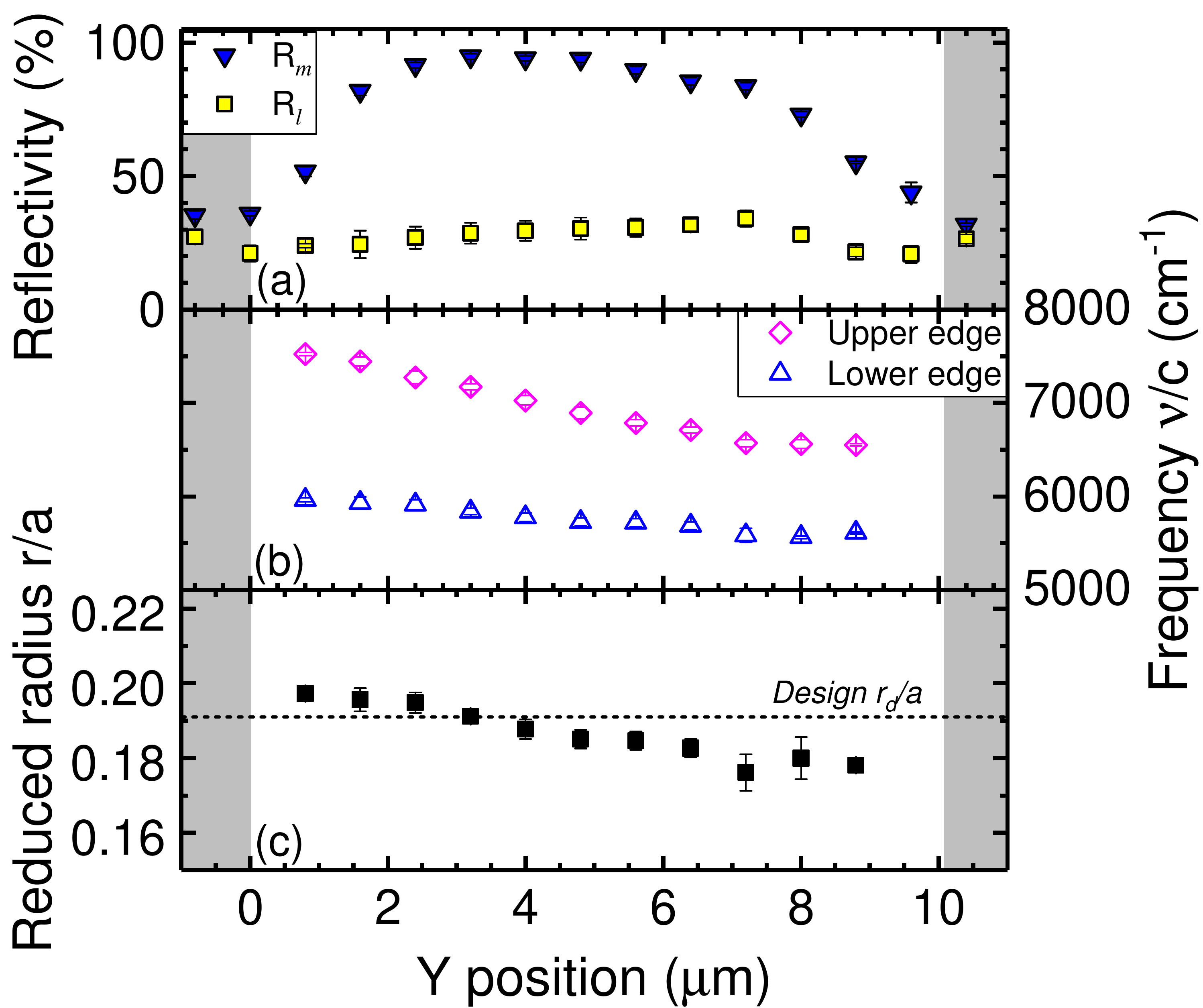}
\caption{Reflectivity measured as a function of Y-position on a crystal with design pore radius $r_d = 130$~nm (or $r_d/a = 0.191$) on silicon beam \textbf{A}, measured with $p$-polarized light. 
(a) Maximum peak reflectivity $(r_m)$ and minimum reflectivity below the stopband $(r_l)$.
(b) Upper edges (magenta diamonds) and lower edges (blue triangles) of the stopband obtained from the half heights of the reflectivity peaks.
(c) Relative radii $r/a$ derived by comparing the lower edge of the stopband with data shown in Fig.~\ref{Fig:bandstruct}(b). 
The grey areas at $Y < 0~\mu$m and $Y > 10~\mu$m indicate bulk silicon outside the crystal with a constant reflectivity near $31\%$. 
}
\label{fig:Yscan}
\end{figure}

Next, we collect reflectivity spectra while scanning the focus across the crystal surface. 
Since we then effectively scan the pore radius $r$, we expect to scan the stopband in response. 
As an example, Figure~\ref{fig:Yscan} shows the results of a $Y$-scan across one of our crystals with design pore radius $r_d = 130$~nm ($r_d/a = 0.191$) on silicon beam \textbf{A}, measured with $p$-polarized light. 
While scanning the $Y$-position, a slight excursion occurred in the $X$-direction from $X = 2.8~\mu$m to $3.2~\mu$m due to imperfect alignment of the silicon beam axis with the vertical axis of the translation stage. 
From each collected spectrum, we derive the peak reflectivity $R_m$ and the minimum reflectivity below the stopband $R_l$ as shown in Figure~\ref{fig:Yscan}(a).
Inside the crystal there is substantial difference between $R_m$ (up to $R_m = 94.8 \%$) and $R_l$, hence the crystal's reflectivity peaks are well-developed. 
Near the crystal edges ($Y = 0~\mu$m and $ 10~\mu$m) the difference between $R_m$ and $R_l$ rapidly decreases and both tend to about $31 \%$ since the focused light here is reflected by bulk silicon. 

Figure~\ref{fig:Yscan}(b) shows the edges of the measured stopband as a function of $Y$-position. 
Between $Y = 0~\mu$m and $ 10~\mu$m the lower edge shifts down from $5950$ to $5550$~cm$^{-1}$ and the upper edge shifts down from $7550$ to $6550$~cm$^{-1}$. 
In other words, both the center frequency of the stopband and its width decrease with increasing $Y$ as a result of the variation of the pore radii with position. 
The redshift of the stopband frequencies is mostly caused by the small excursion along $X$, since the radius of the $X$-directed pores decreases with increasing $X$. 

By comparing the measured lower edges in Figure~\ref{fig:Yscan}(b) with the theoretical gap maps shown in Fig.~\ref{Fig:bandstruct}(b), we derive the local pore radius $r (\vec{\mathbf{r}})$ in the crystal that is plotted versus $Y$-position in Figure~\ref{fig:Yscan}(c). 
The resulting $r (\vec{\mathbf{r}})/a $ is seen to vary from $0.197$ to $0.176$ about the design pore radius  $r_d/a = 0.191$. 
Therefore, we can now combine all position-dependent data to make maps of stopband centers and stopband widths as a function of the pore radius. 

\subsection{Probing the 3D photonic band gap}\label{sec:band-gap}

\begin{figure}[ht]
\centering
\includegraphics[width=80mm]{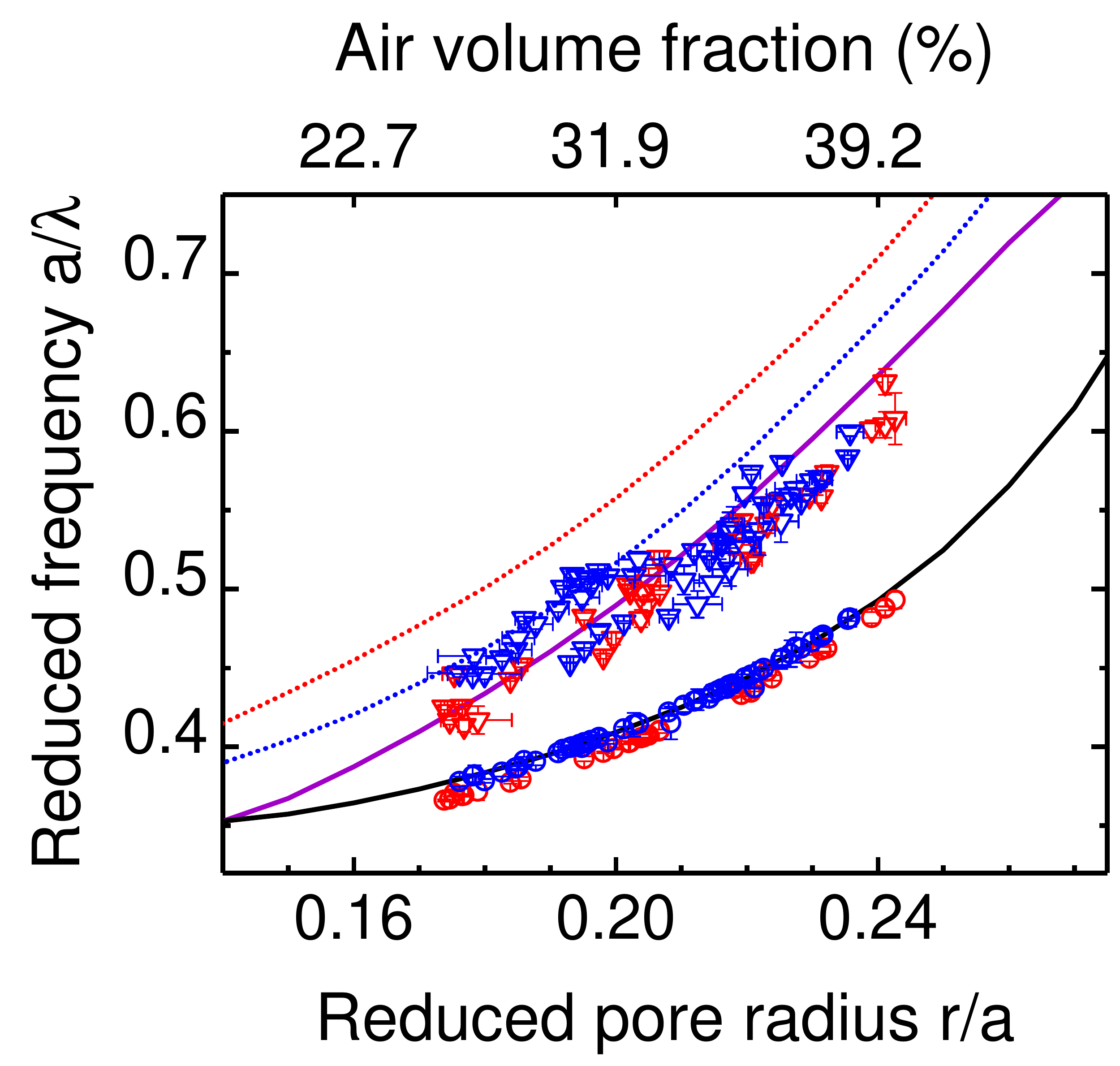}
\caption{Evolution of the stopband edges versus pore radius. 
The red and blue triangles represent upper edge of the stopband for $s$ and $p$-polarized light respectively.
The red and blue circles represent the lower edge of the stopband for $s$ and $p$ polarized light.
The stopband edges are inferred from the reflectivity peak measured on 11 crystals on the Si beam \textbf{A}. 
The solid lines indicate the edges of the photonic band gap.
The upper edge of the $\Gamma Z$ stop gap for $s$ and $p$ polarized light are plotted as the red and blue dotted curves, respectively. 
For both polarizations, the experimental data agree well with the 3D photonic band gap edge.
}
\label{fig:bandedge_rad}
\end{figure}
We have applied the procedures described in sections~\ref{sec:position_dependent_stopbands} and~\ref{sec:stopbands} to reflectivity measured on many crystals on beam \textbf{A}. 
We also took multiple measurements along the $Y$-direction on two crystals to verify the consistency of all observations.
From all collected reflectivity spectra, both $s$ and $p$ polarized, the lower and upper stopband edges are extracted, and are mapped as a function of $r/a$ in Figure~\ref{fig:bandedge_rad}.
The lower edge data form a continuous trace from reduced frequency $a/\lambda = 0.38$ at $r/a = 0.17$ to  $a/\lambda = 0.50$ at $r/a = 0.245$. 
The data match well with the theory, which is obvious since we used the lower edge to estimate $r/a$ from the measured spectra. 
The upper edge data form a continuous trace from reduced frequency $a/\lambda = 0.42$ at $r/a = 0.17$ to $a/\lambda = 0.64$ at $r/a = 0.245$. 
It is remarkable that the upper edge data for both $s$ and $p$-polarized light mutually agree very well, especially for pores radii $r/a > 0.21$. 
This observation implies that the measured stopband is rather representative of the photonic band gap that is polarization insensitive, as opposed to a directional stop gap that is polarization sensitive.
In comparison to theory, at pore radii $r/a < 0.21$, the upper edges are in between the theoretical upper edges of the band gap and the $p$-polarized edge of the directional stop gap. 
At larger radii $(r/a > 0.21)$ all measured upper edge data are near the theoretical upper band gap edge and differ from the stopband edges. 
This observation adds support to the notion that the structure-dependent stopbands represent the 3D photonic band gap, rather than a directional stop gap. 

\begin{figure}[ht]
\centering
\includegraphics[width=80mm]{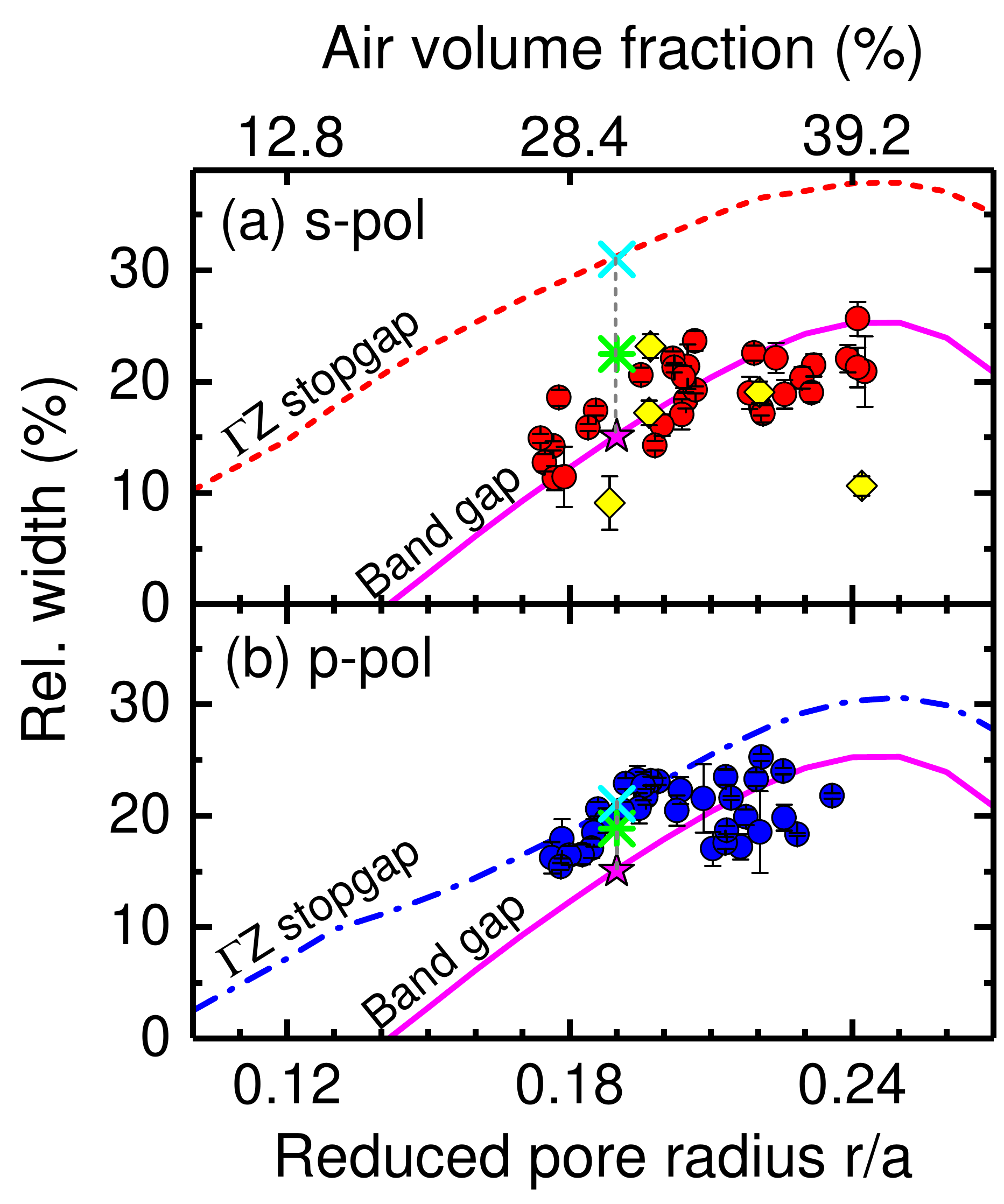}
\caption{Measured relative stopband width (gap width to midgap, $\Delta \omega / \omega_{c}$) versus reduced pore radii $r/a$ (circles). 
The $r/a$ values are estimated from the lower edge of the stopband, see Fig.~\ref{Fig:bandstruct}(b). 
(a) $s$-polarized data for beam \textbf{A} (red circles), and for beam \textbf{B} (yellow diamonds), (b) p-polarized data for beam \textbf{A} (blue circles). 
The cyan crosses, green asterisks, and magenta stars are numerical results for angle-averaged stopband, normal incidence, and band gap at $r/a=0.19$, respectively~\cite{Devashish2017PRB}.
The dashed red and dash dotted blue curves represent the width of the $\Gamma Z$ stop gap obtained from band structures for an infinite crystal for $s$ and $p$ polarized light, respectively.
The magenta solid curve is the 3D photonic band gap from band structures for an infinite crystal. 
}
\label{fig:rel-width_vs_radius}
\end{figure}

To refine our reasoning, we plot in Figures~\ref{fig:rel-width_vs_radius}(a,b) the relative stopband width (gap to mid-gap ratio) as a function of the reduced pore radius $r/a$ as derived from the lower edges. 
The large number of data in Figure~\ref{fig:rel-width_vs_radius}(a) for Si beam \textbf{A} show that the width of the $s$-polarized stopband increases up to $r/a = 0.2$ before more or less saturating up to $r/a = 0.24$. 
The $s$-polarized data for Si beam \textbf{B} agree well with the data for beam \textbf{A} except for an outlier at $r/a = 0.24$. 
All data are close to the theoretical prediction for the width of the 3D photonic band gap and lie distinctly below the theoretical width of the stop gap. 
Figure~\ref{fig:rel-width_vs_radius}(a) also shows results of $s$-polarized reflectivity simulated for a \textit{finite} inverse woodpile crystal $r/a=0.19$~\cite{Devashish2017PRB}, namely of a directional stopband, of an angle-averaged stopband (for a range of angles relevant for a reflecting objective with $NA = 0.65$), and of an omnidirectional band gap. 
With increasing aperture, the simulated stopband becomes narrower. 
From the comparison, it is apparent that our data match best with the width of the 3D photonic band gap. 

Figure~\ref{fig:rel-width_vs_radius}(b) shows the $p$-polarized stopband widths versus pore radius. 
At pore radii $r/a < 0.21$, the stopband widths are in between the theoretical bandwidths of either the directional stopgap or the omnidirectional band gap. 
At larger radii $(r/a > 0.21)$, the measured stopband widths match better with the theoretical width of the band gap than with the stop gap width. 
From $p$-polarized finite-crystal simulations done at $r/a=0.19$~\cite{Devashish2017PRB}, we learn that the bandwidths of the directional stop gap, of the angle-averaged stopgap, and of the band gap are near to each other, hence it is difficult given the variations in our data to discriminate between either feature. 
Considering the $s$ and $p$-polarized stopband widths jointly, we again find a much better agreement with the 3D photonic band gap than with the directional stop gap.

\begin{figure}[ht]
\includegraphics[width=80mm]{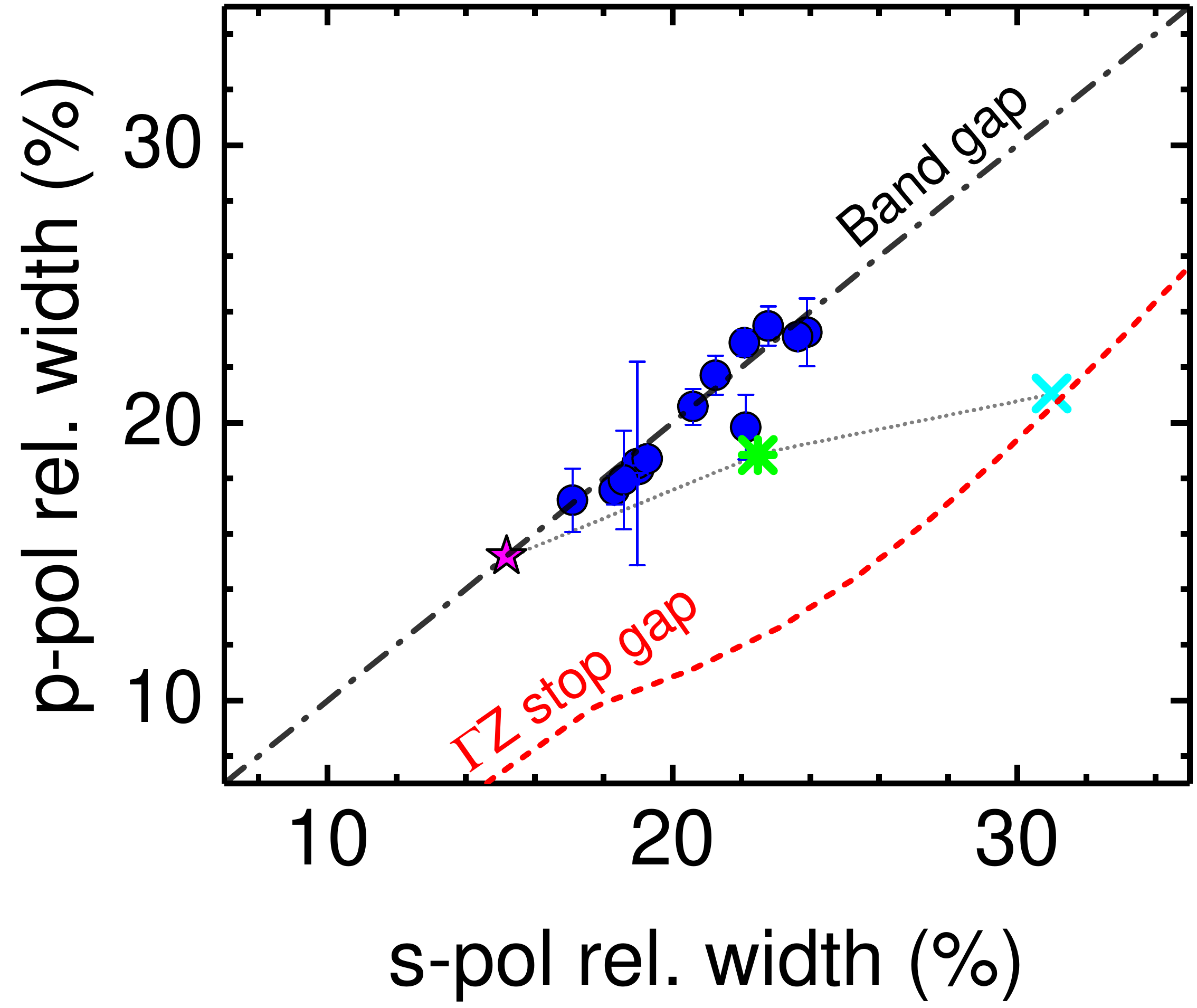}
\centering
\caption{Relative stopband width for p-polarization versus relative stopband width for s-polarization measured at the same position on crystals with a range of pore radii $r/a$ (blue circles). 
Black dashed-dotted line is the infinite-crystal theory result for the 3D photonic band gap, and red dashed curve the one for the $\Gamma Z$ stop gap. 
The cyan crosses and green asterisks are numerical results for angle-averaged stopband and normal incidence for $r/a=0.19$, respectively, and the magenta star is the band gap width simulated for a finite-thickness crystal with $r/a=0.19$~\cite{Devashish2017PRB}, that are connected by the gray dotted line as a guide to the eye. 
}
\label{fig:ppol-width_vs_spol-width}
\end{figure}
The conclusions from Figures~\ref{fig:bandedge_rad} and~\ref{fig:rel-width_vs_radius} are based on the agreement between measurements on one hand, and simulations and theory on the other hand. 
The latter invoke an idealized structural model, for instance, pores as infinite perfect cylinders which neglect pore tapering. 
Therefore, to find a criterion that is indeed free of theoretical or numerical modeling, we make a parametric plot of the width of the $p$-polarized stopband versus the width of the $s$-polarized stopband, as shown in Figure~\ref{fig:ppol-width_vs_spol-width}.
In order to avoid systematic errors due to the position-dependence of the stopbands, we select data where both polarizations were measured on the same position on a crystal. 
For $s$-polarized stopband widths between $\Delta \omega/\omega_c = 17\%$ and $24\%$, the corresponding $p$-polarized stopband width increases linearly, and also from $17 \%$ to $24\%$. 
Such a linear increase is obviously expected for a 3D photonic band gap, even without detailed modeling, since a 3D band gap obviously entails a forbidden gap for both polarizations simultaneously~\cite{Joannopoulos2008PhotonicLight}. 
In case of the alternative hypothesis that the stopbands correspond to directional $\Gamma Z$ stop gaps, the trend would be nonlinear and clearly different from the diagonal. 
Since this trend obviously does not match with our data, we can safely reject this hypothesis. 

For comparison, the computer simulations on a finite-size crystal (with  $r/a = 0.19$) in Ref.~\cite{Devashish2017PRB} agree with the theory both for the omnidirectional photonic band gap and for the directional stopgap, where the former matches very well with our observations and the latter does not. 
The simulations have also been done for a numerical aperture comparable to a reflecting objective (as in Ref.~\cite{Huisman2011PRB}), and this result is somewhat lower than our observations, which indicates that for a smaller NA than studied here the measured stopband is not representative of the band gap. 
Conversely, the numerical aperture $NA = 0.85$ used here and the correspondingly large overall solid angle of $3.8\pi$ sr is apparently sufficient to probe the omnidirectional photonic band gap. 

\section{Discussion}\label{sec:discussion}

\begin{figure}[ht]
\includegraphics[width=80mm]{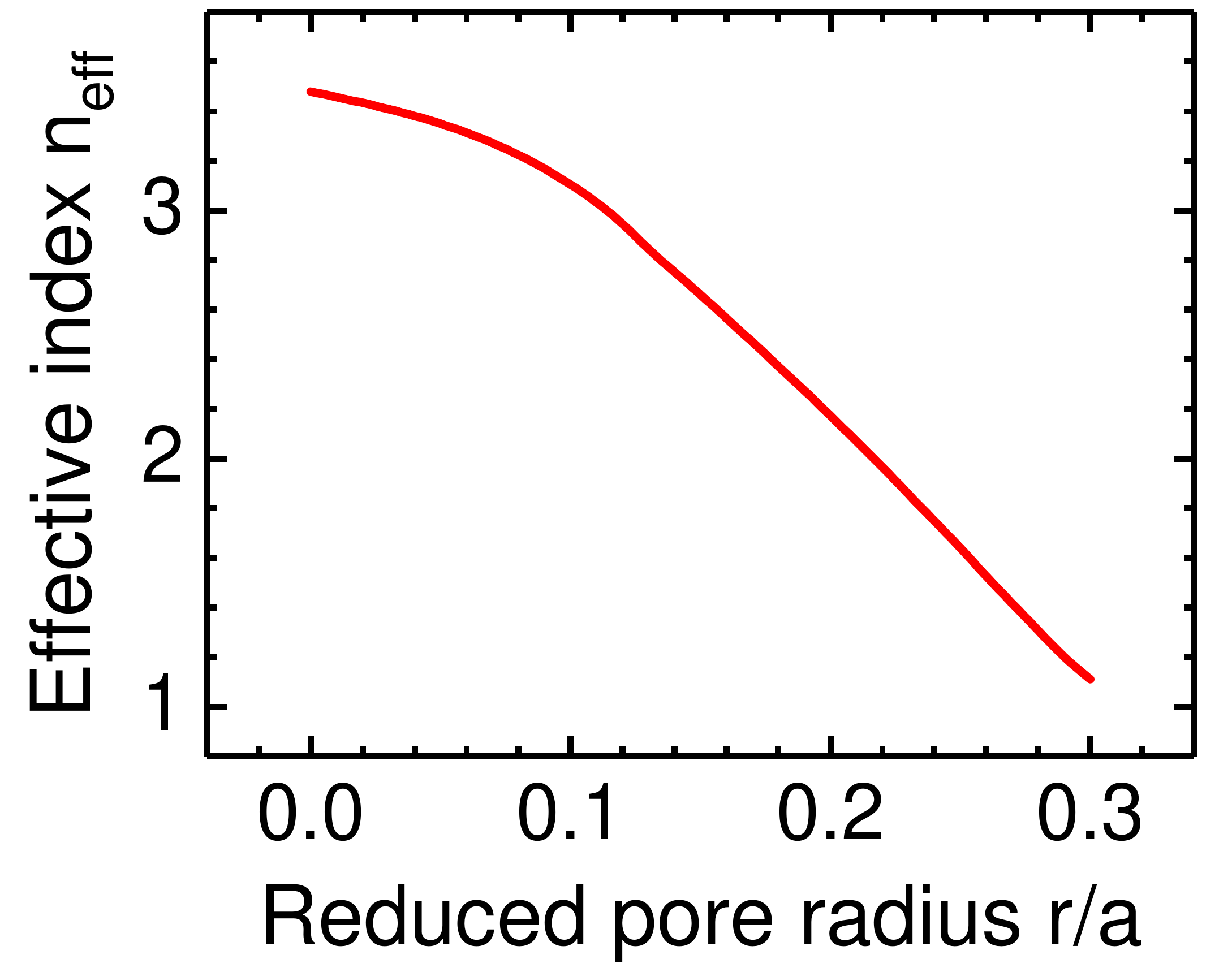}
\centering
\caption{Effective refractive index of inverse woodpile photonic crystals made of silicon as a function of the pore reduced pore radius $r/a$, obtained from the slope of the band structures in the limit of zero frequency.} 
\label{fig:neff_vs_pore-radius}
\end{figure}

So far, we discussed the stopbands versus the radii of the pores that are specific to the inverse woodpile structure studied here~\cite{Ho1994SSC}. 
In order to generalize our results to other classes of photonic band gap crystals, such as inverse opals, direct woodpiles, and even non-periodic ones~\cite{Muller2017Optica}, it is useful to realize that a varying pore size corresponds to the tuning of the filling fraction and thus the tuning of the effective refractive index~\cite{Datta1993PRB}, both of which pertain to all other classes of photonic band gap structures. 
As is shown in Figure~\ref{fig:neff_vs_pore-radius}, the effective index of our crystals obtained from the band structures in the limit of zero frequency - is tuned from $3.5$ (silicon) to $1.0$ (air) by varying the pore size from $r/a = 0.0$ to a little over $0.3$. 
Both the filling fraction and the effective index are readily generalized to other 3D photonic band gap crystals. 
For instance, in inverse opals the filling fraction of the high-index backbone is known to vary with preparation conditions~\cite{Wijnhoven2001CM}, hence this can be used as a tuning knob. 
In direct woodpile crystals, the filling fraction is notably tuned by varying the width of the high-index nanorods~\cite{Noda2000Science, Tajiri2019Optica}, and similarly in hyperuniform structures~\cite{Muller2017Optica}. 
It is therefore that the top abscissae in Figures~\ref{fig:rel-width_vs_radius} and~\ref{fig:ppol-width_vs_spol-width} have been generalized to the effective refractive index. 
Therefore, the stopband width versus the effective index (as in Fig.~\ref{fig:rel-width_vs_radius}) or the $p$-polarized stopband width versus the $s$-polarized one also pertain as probes to other classes of band gap structures, and thus serve as experimental probes of the 3D photonic band gap in such other structures. 


It is generally agreed that the fabrication of 3D nanostructures necessary for photonic band gap physics is fairly challenging~\cite{Wijnhoven1998Science, Noda2000Science, Blanco2000Nature, Vlasov2001N}. 
Since the detailed 3D nanostructure critically determines the band gap functionality, it is important to have a non-destructive verification on the functionality. 
We propose that the practical probe methods presented here fill this gap by providing relatively fast feedback on a newly fabricated nanostructure. 
In a fundamentally holistic approach, one would not only verify the functionality but also the 3D structure since the latter usually serves to improve the understanding of the functionality, especially in ubiquitous situations where the function differs from the designed one. 
While studying the detailed 3D structure of a nanostructure is highly non-trivial, successful methods have been reported using X-ray techniques, notably small-angle X-ray scattering~\cite{Wijnhoven2001CM}, X-ray ptychography~\cite{Furlan2018ApplMatToday}, or traceless X-ray tomography~\cite{Grishina2018Arxiv}. 

We expect that a practical probe of 3D photonic band gaps will boost their applications in several innovative fields. 
For instance, recent efforts by the Tokyo and Kyoto teams have demonstrated the use of 3D photonic band gap crystals as platforms for 3D photonic integrated circuits~\cite{Tajiri2019Optica, Ishizaki2013NatPhot}. 
In the field of photovoltaics that is of considerable societal interest, the use of 3D photonic band gap crystals is increasingly studied to enhance the collection efficiency by means of various kinds of photon management~\cite{Bermel2007OE, Wehrspohn2012JO, Devashish2019PRB}. 
It is an essential feature of a 3D photonic band gap crystal to have a gap in the density of states, which in turn corresponds to the density of vacuum fluctuations. 
Therefore, quantum devices embedded inside a 3D band gap crystal are effectively shielded from quantum noise~\cite{Clerk2010RMP}, including quantum gates that manipulate qubits for quantum information processing.

\section{Conclusion}\label{sec:conclusion}
We present a purely experimental probe of the 3D band gap in real three-dimensional (3D) photonic crystals, without the need for theoretical or numerical modeling that invoke idealized and even infinite photonic crystals. 
As an exemplary structure, we study 3D inverse woodpile crystals made from silicon. 
We collected position and polarization-resolved reflectivity spectra of multiple crystals with different design parameters with a large numerical aperture and observed intense reflectivity peaks with maxima exceeding $90\%$ corresponding to the stopbands, typical of high-quality crystals. 
We track the stopband width versus pore radius, which agrees much better with the predicted 3D photonic band gap than with a directional stop gap. 
A parametric plot of s-polarized versus p-polarized stopband width is nearly a straight line, in agreement with the 3D band gap and at variance with the directional stop gap. 
Such a practical probe provides fast feedback on the advanced nanofabrication required for 3D photonic crystals and stimulates practical applications of band gaps in 3D silicon nanophotonics and photonic integrated circuits, photovoltaics, cavity QED, and quantum information processing. 

\section*{Funding}
NWO-TTW Perspectief Program `Free Form Scattering Optics', NWO-FOM program `Stirring of Light!', MESA+ section `Applied Nanophotonics (ANP)'. 

\section*{Acknowledgments}
We thank Rajesh Nair, Simon Huisman, and Devashish for help with the photonic band structure calculations and Emre Yuce for early contributions in building the reflectivity setup. 

\bibliography{3D_pbg_probe.bib}

\begin{thebibliography}{75}%
\makeatletter
\providecommand \@ifxundefined [1]{%
 \@ifx{#1\undefined}
}%
\providecommand \@ifnum [1]{%
 \ifnum #1\expandafter \@firstoftwo
 \else \expandafter \@secondoftwo
 \fi
}%
\providecommand \@ifx [1]{%
 \ifx #1\expandafter \@firstoftwo
 \else \expandafter \@secondoftwo
 \fi
}%
\providecommand \natexlab [1]{#1}%
\providecommand \enquote  [1]{``#1''}%
\providecommand \bibnamefont  [1]{#1}%
\providecommand \bibfnamefont [1]{#1}%
\providecommand \citenamefont [1]{#1}%
\providecommand \href@noop [0]{\@secondoftwo}%
\providecommand \href [0]{\begingroup \@sanitize@url \@href}%
\providecommand \@href[1]{\@@startlink{#1}\@@href}%
\providecommand \@@href[1]{\endgroup#1\@@endlink}%
\providecommand \@sanitize@url [0]{\catcode `\\12\catcode `\$12\catcode
  `\&12\catcode `\#12\catcode `\^12\catcode `\_12\catcode `\%12\relax}%
\providecommand \@@startlink[1]{}%
\providecommand \@@endlink[0]{}%
\providecommand \url  [0]{\begingroup\@sanitize@url \@url }%
\providecommand \@url [1]{\endgroup\@href {#1}{\urlprefix }}%
\providecommand \urlprefix  [0]{URL }%
\providecommand \Eprint [0]{\href }%
\providecommand \doibase [0]{http://dx.doi.org/}%
\providecommand \selectlanguage [0]{\@gobble}%
\providecommand \bibinfo  [0]{\@secondoftwo}%
\providecommand \bibfield  [0]{\@secondoftwo}%
\providecommand \translation [1]{[#1]}%
\providecommand \BibitemOpen [0]{}%
\providecommand \bibitemStop [0]{}%
\providecommand \bibitemNoStop [0]{.\EOS\space}%
\providecommand \EOS [0]{\spacefactor3000\relax}%
\providecommand \BibitemShut  [1]{\csname bibitem#1\endcsname}%
\let\auto@bib@innerbib\@empty
\bibitem [{\citenamefont {Novotny}\ and\ \citenamefont
  {Hecht}(2006)}]{Novotny2006book}%
  \BibitemOpen
  \bibfield  {author} {\bibinfo {author} {\bibfnamefont {L.}~\bibnamefont
  {Novotny}}\ and\ \bibinfo {author} {\bibfnamefont {B.}~\bibnamefont
  {Hecht}},\ }\href@noop {} {\emph {\bibinfo {title} {Principles of
  Nano-Optics}}}\ (\bibinfo  {publisher} {Cambridge University Press,
  Cambridge},\ \bibinfo {year} {2006})\BibitemShut {NoStop}%
\bibitem [{\citenamefont {Joannopoulos}\ \emph {et~al.}(2008)\citenamefont
  {Joannopoulos}, \citenamefont {Johnson}, \citenamefont {Winn},\ and\
  \citenamefont {Meade}}]{Joannopoulos2008PhotonicLight}%
  \BibitemOpen
  \bibfield  {author} {\bibinfo {author} {\bibfnamefont {J.~D.}\ \bibnamefont
  {Joannopoulos}}, \bibinfo {author} {\bibfnamefont {S.}~\bibnamefont
  {Johnson}}, \bibinfo {author} {\bibfnamefont {J.~N.}\ \bibnamefont {Winn}}, \
  and\ \bibinfo {author} {\bibfnamefont {R.~D.}\ \bibnamefont {Meade}},\ }\href
  {\doibase 10.1063/1.1586781} {\emph {\bibinfo {title} {Photonic crystals:
  molding the flow of light}}}\ (\bibinfo  {publisher} {Princeton University
  Press, Princeton NJ},\ \bibinfo {year} {2008})\BibitemShut {NoStop}%
\bibitem [{\citenamefont {Lourtioz}\ \emph {et~al.}(2008)\citenamefont
  {Lourtioz}, \citenamefont {Benisty}, \citenamefont {Berger}, \citenamefont
  {G{\'{e}}rard}, \citenamefont {Maystre},\ and\ \citenamefont
  {Tchelnokov}}]{Lourtioz2008book}%
  \BibitemOpen
  \bibfield  {author} {\bibinfo {author} {\bibfnamefont {J.-M.}\ \bibnamefont
  {Lourtioz}}, \bibinfo {author} {\bibfnamefont {H.}~\bibnamefont {Benisty}},
  \bibinfo {author} {\bibfnamefont {V.}~\bibnamefont {Berger}}, \bibinfo
  {author} {\bibfnamefont {J.-M.}\ \bibnamefont {G{\'{e}}rard}}, \bibinfo
  {author} {\bibfnamefont {D.}~\bibnamefont {Maystre}}, \ and\ \bibinfo
  {author} {\bibfnamefont {A.}~\bibnamefont {Tchelnokov}},\ }\href@noop {}
  {\emph {\bibinfo {title} {Photonic Crystals}}}\ (\bibinfo  {publisher}
  {Springer Verlag, Heidelberg-Berlin},\ \bibinfo {year} {2008})\BibitemShut
  {NoStop}%
\bibitem [{\citenamefont {Noginov}\ \emph {et~al.}(2009)\citenamefont
  {Noginov}, \citenamefont {Dewar}, \citenamefont {McCall},\ and\ \citenamefont
  {Zheludev}}]{Noginov2009book}%
  \BibitemOpen
  \bibinfo {editor} {\bibfnamefont {M.~A.}\ \bibnamefont {Noginov}}, \bibinfo
  {editor} {\bibfnamefont {G.}~\bibnamefont {Dewar}}, \bibinfo {editor}
  {\bibfnamefont {M.~W.}\ \bibnamefont {McCall}}, \ and\ \bibinfo {editor}
  {\bibfnamefont {N.~I.}\ \bibnamefont {Zheludev}},\ eds.,\ \href@noop {}
  {\emph {\bibinfo {title} {{Tutorials in Complex Photonic Media}}}}\ (\bibinfo
   {publisher} {Cambridge University Press, Cambridge},\ \bibinfo {year}
  {2009})\BibitemShut {NoStop}%
\bibitem [{\citenamefont {Ghulinyan}\ and\ \citenamefont
  {Pavesi}(2014)}]{Ghulinyan2015book}%
  \BibitemOpen
  \bibinfo {editor} {\bibfnamefont {M.}~\bibnamefont {Ghulinyan}}\ and\
  \bibinfo {editor} {\bibfnamefont {L.}~\bibnamefont {Pavesi}},\ eds.,\
  \href@noop {} {\emph {\bibinfo {title} {Light Localisation and lasing: Random
  and quasi-random photonic structures}}}\ (\bibinfo  {publisher} {Cambridge
  University Press, Cambridge},\ \bibinfo {year} {2014})\BibitemShut {NoStop}%
\bibitem [{\citenamefont {Ashcroft}\ and\ \citenamefont
  {Mermin}(1976)}]{Ashcroft1976book}%
  \BibitemOpen
  \bibfield  {author} {\bibinfo {author} {\bibfnamefont {N.~W.}\ \bibnamefont
  {Ashcroft}}\ and\ \bibinfo {author} {\bibfnamefont {N.~D.}\ \bibnamefont
  {Mermin}},\ }\href@noop {} {\emph {\bibinfo {title} {Solid State Physics}}}\
  (\bibinfo  {publisher} {Holt, Rinehart and Winston, New York NY},\ \bibinfo
  {year} {1976})\BibitemShut {NoStop}%
\bibitem [{\citenamefont {Economou}(2010)}]{Economou2010book}%
  \BibitemOpen
  \bibfield  {author} {\bibinfo {author} {\bibfnamefont {E.~N.}\ \bibnamefont
  {Economou}},\ }\href@noop {} {\emph {\bibinfo {title} {The Physics of
  Solids}}}\ (\bibinfo  {publisher} {Springer-Verlag, Berlin Heidelberg},\
  \bibinfo {year} {2010})\BibitemShut {NoStop}%
\bibitem [{\citenamefont {van Driel}\ and\ \citenamefont
  {Vos}(2000)}]{vanDriel2000PRB}%
  \BibitemOpen
  \bibfield  {author} {\bibinfo {author} {\bibfnamefont {H.~M.}\ \bibnamefont
  {van Driel}}\ and\ \bibinfo {author} {\bibfnamefont {W.~L.}\ \bibnamefont
  {Vos}},\ }\href@noop {} {\bibfield  {journal} {\bibinfo  {journal} {Phys.
  Rev. B}\ }\textbf {\bibinfo {volume} {62}},\ \bibinfo {pages} {9872}
  (\bibinfo {year} {2000})}\BibitemShut {NoStop}%
\bibitem [{\citenamefont {Vos}\ and\ \citenamefont {van
  Driel}(2000)}]{Vos2000PLA}%
  \BibitemOpen
  \bibfield  {author} {\bibinfo {author} {\bibfnamefont {W.~L.}\ \bibnamefont
  {Vos}}\ and\ \bibinfo {author} {\bibfnamefont {H.~M.}\ \bibnamefont {van
  Driel}},\ }\href@noop {} {\bibfield  {journal} {\bibinfo  {journal} {Phys.
  Lett. A}\ }\textbf {\bibinfo {volume} {272}},\ \bibinfo {pages} {101}
  (\bibinfo {year} {2000})}\BibitemShut {NoStop}%
\bibitem [{\citenamefont {Romanov}\ \emph {et~al.}(2001)\citenamefont
  {Romanov}, \citenamefont {Maka}, \citenamefont {Sotomayor~Torres},
  \citenamefont {Müller}, \citenamefont {Zentel}, \citenamefont {Cassagne},
  \citenamefont {Manzanares-Martinez},\ and\ \citenamefont
  {Jouanin}}]{Romanov2001PRE}%
  \BibitemOpen
  \bibfield  {author} {\bibinfo {author} {\bibfnamefont {S.~G.}\ \bibnamefont
  {Romanov}}, \bibinfo {author} {\bibfnamefont {T.}~\bibnamefont {Maka}},
  \bibinfo {author} {\bibfnamefont {C.~M.}\ \bibnamefont {Sotomayor~Torres}},
  \bibinfo {author} {\bibfnamefont {M.}~\bibnamefont {Müller}}, \bibinfo
  {author} {\bibfnamefont {R.}~\bibnamefont {Zentel}}, \bibinfo {author}
  {\bibfnamefont {D.}~\bibnamefont {Cassagne}}, \bibinfo {author}
  {\bibfnamefont {J.}~\bibnamefont {Manzanares-Martinez}}, \ and\ \bibinfo
  {author} {\bibfnamefont {C.}~\bibnamefont {Jouanin}},\ }\href@noop {}
  {\bibfield  {journal} {\bibinfo  {journal} {Phys. Rev. E}\ }\textbf {\bibinfo
  {volume} {63}},\ \bibinfo {pages} {056603} (\bibinfo {year}
  {2001})}\BibitemShut {NoStop}%
\bibitem [{\citenamefont {Bykov}(1972)}]{Bykov1972JETP}%
  \BibitemOpen
  \bibfield  {author} {\bibinfo {author} {\bibfnamefont {V.~P.}\ \bibnamefont
  {Bykov}},\ }\href@noop {} {\bibfield  {journal} {\bibinfo  {journal} {Sov.
  Phys. JETP}\ }\textbf {\bibinfo {volume} {35}},\ \bibinfo {pages} {269}
  (\bibinfo {year} {1972})}\BibitemShut {NoStop}%
\bibitem [{\citenamefont {Yablonovitch}(1987)}]{Yablonovitch1987PRL}%
  \BibitemOpen
  \bibfield  {author} {\bibinfo {author} {\bibfnamefont {E.}~\bibnamefont
  {Yablonovitch}},\ }\href {\doibase 10.1103/PhysRevLett.58.2059} {\bibfield
  {journal} {\bibinfo  {journal} {Physical Review Letters}\ }\textbf {\bibinfo
  {volume} {58}},\ \bibinfo {pages} {2059} (\bibinfo {year}
  {1987})}\BibitemShut {NoStop}%
\bibitem [{\citenamefont {John}\ and\ \citenamefont
  {Wang}(1990)}]{John1990PRL}%
  \BibitemOpen
  \bibfield  {author} {\bibinfo {author} {\bibfnamefont {S.}~\bibnamefont
  {John}}\ and\ \bibinfo {author} {\bibfnamefont {J.}~\bibnamefont {Wang}},\
  }\href {\doibase 10.1103/PhysRevLett.64.2418} {\bibfield  {journal} {\bibinfo
   {journal} {Physical Review Letters}\ }\textbf {\bibinfo {volume} {64}},\
  \bibinfo {pages} {2418} (\bibinfo {year} {1990})}\BibitemShut {NoStop}%
\bibitem [{\citenamefont {Vos}\ and\ \citenamefont
  {Woldering}(2015)}]{Vos2015CavityCrystals}%
  \BibitemOpen
  \bibfield  {author} {\bibinfo {author} {\bibfnamefont {W.~L.}\ \bibnamefont
  {Vos}}\ and\ \bibinfo {author} {\bibfnamefont {L.~A.}\ \bibnamefont
  {Woldering}},\ }in\ \href@noop {} {\emph {\bibinfo {booktitle} {Light
  Localisation and Lasing}}},\ \bibinfo {editor} {edited by\ \bibinfo {editor}
  {\bibfnamefont {M.}~\bibnamefont {Ghulinyan}}\ and\ \bibinfo {editor}
  {\bibfnamefont {L.}~\bibnamefont {Pavesi}}}\ (\bibinfo  {publisher}
  {Cambridge University Press, Cambridge},\ \bibinfo {year} {2015})\ p.\
  \bibinfo {pages} {180}\BibitemShut {NoStop}%
\bibitem [{\citenamefont {Smith}\ \emph {et~al.}(1998)\citenamefont {Smith},
  \citenamefont {Kesler},\ and\ \citenamefont {Maloney}}]{Smith1998MOTL}%
  \BibitemOpen
  \bibfield  {author} {\bibinfo {author} {\bibfnamefont {G.~S.}\ \bibnamefont
  {Smith}}, \bibinfo {author} {\bibfnamefont {M.~P.}\ \bibnamefont {Kesler}}, \
  and\ \bibinfo {author} {\bibfnamefont {J.~G.}\ \bibnamefont {Maloney}},\
  }\href@noop {} {\bibfield  {journal} {\bibinfo  {journal} {Microwave and
  Optical Technology Letters}\ }\textbf {\bibinfo {volume} {21}},\ \bibinfo
  {pages} {191} (\bibinfo {year} {1998})}\BibitemShut {NoStop}%
\bibitem [{\citenamefont {Bermel}\ \emph {et~al.}(2007)\citenamefont {Bermel},
  \citenamefont {Luo}, \citenamefont {Zeng}, \citenamefont {Kimerling},\ and\
  \citenamefont {Joannopoulos}}]{Bermel2007OE}%
  \BibitemOpen
  \bibfield  {author} {\bibinfo {author} {\bibfnamefont {P.}~\bibnamefont
  {Bermel}}, \bibinfo {author} {\bibfnamefont {C.}~\bibnamefont {Luo}},
  \bibinfo {author} {\bibfnamefont {L.}~\bibnamefont {Zeng}}, \bibinfo {author}
  {\bibfnamefont {L.~C.}\ \bibnamefont {Kimerling}}, \ and\ \bibinfo {author}
  {\bibfnamefont {J.~D.}\ \bibnamefont {Joannopoulos}},\ }\href@noop {}
  {\bibfield  {journal} {\bibinfo  {journal} {Optics Express}\ }\textbf
  {\bibinfo {volume} {15}},\ \bibinfo {pages} {16986} (\bibinfo {year}
  {2007})}\BibitemShut {NoStop}%
\bibitem [{\citenamefont {Wehrspohn}\ and\ \citenamefont
  {{\"{U}}pping}(2012)}]{Wehrspohn2012JO}%
  \BibitemOpen
  \bibfield  {author} {\bibinfo {author} {\bibfnamefont {R.~B.}\ \bibnamefont
  {Wehrspohn}}\ and\ \bibinfo {author} {\bibfnamefont {J.}~\bibnamefont
  {{\"{U}}pping}},\ }\href@noop {} {\bibfield  {journal} {\bibinfo  {journal}
  {Journal of Optics}\ }\textbf {\bibinfo {volume} {14}},\ \bibinfo {pages}
  {024003} (\bibinfo {year} {2012})}\BibitemShut {NoStop}%
\bibitem [{\citenamefont {Koenderink}\ \emph {et~al.}(2015)\citenamefont
  {Koenderink}, \citenamefont {Al{\'{u}}},\ and\ \citenamefont
  {Polman}}]{Koenderink2015science}%
  \BibitemOpen
  \bibfield  {author} {\bibinfo {author} {\bibfnamefont {A.~F.}\ \bibnamefont
  {Koenderink}}, \bibinfo {author} {\bibfnamefont {A.}~\bibnamefont
  {Al{\'{u}}}}, \ and\ \bibinfo {author} {\bibfnamefont {A.}~\bibnamefont
  {Polman}},\ }\href {\doibase 10.1126/science.1261243} {\bibfield  {journal}
  {\bibinfo  {journal} {Science}\ }\textbf {\bibinfo {volume} {348}},\ \bibinfo
  {pages} {516} (\bibinfo {year} {2015})}\BibitemShut {NoStop}%
\bibitem [{\citenamefont {David}\ \emph {et~al.}(2012)\citenamefont {David},
  \citenamefont {Benisty},\ and\ \citenamefont {Weisbuch}}]{David2012RPP}%
  \BibitemOpen
  \bibfield  {author} {\bibinfo {author} {\bibfnamefont {A.}~\bibnamefont
  {David}}, \bibinfo {author} {\bibfnamefont {H.}~\bibnamefont {Benisty}}, \
  and\ \bibinfo {author} {\bibfnamefont {C.}~\bibnamefont {Weisbuch}},\
  }\href@noop {} {\bibfield  {journal} {\bibinfo  {journal} {Rep. Prog. Phys.}\
  }\textbf {\bibinfo {volume} {75}},\ \bibinfo {pages} {126501} (\bibinfo
  {year} {2012})}\BibitemShut {NoStop}%
\bibitem [{\citenamefont {Li}\ and\ \citenamefont {Ho}(2003)}]{Li2003JOSA}%
  \BibitemOpen
  \bibfield  {author} {\bibinfo {author} {\bibfnamefont {Z.-Y.}\ \bibnamefont
  {Li}}\ and\ \bibinfo {author} {\bibfnamefont {K.-M.}\ \bibnamefont {Ho}},\
  }\href@noop {} {\bibfield  {journal} {\bibinfo  {journal} {Journal of Optical
  Society of America B}\ }\textbf {\bibinfo {volume} {20}},\ \bibinfo {pages}
  {801} (\bibinfo {year} {2003})}\BibitemShut {NoStop}%
\bibitem [{\citenamefont {Tajiri}\ \emph {et~al.}(2019)\citenamefont {Tajiri},
  \citenamefont {Takahashi}, \citenamefont {Ota}, \citenamefont {Watanabe},
  \citenamefont {Iwamoto},\ and\ \citenamefont {Arakawa}}]{Tajiri2019Optica}%
  \BibitemOpen
  \bibfield  {author} {\bibinfo {author} {\bibfnamefont {T.}~\bibnamefont
  {Tajiri}}, \bibinfo {author} {\bibfnamefont {S.}~\bibnamefont {Takahashi}},
  \bibinfo {author} {\bibfnamefont {Y.}~\bibnamefont {Ota}}, \bibinfo {author}
  {\bibfnamefont {K.}~\bibnamefont {Watanabe}}, \bibinfo {author}
  {\bibfnamefont {S.}~\bibnamefont {Iwamoto}}, \ and\ \bibinfo {author}
  {\bibfnamefont {Y.}~\bibnamefont {Arakawa}},\ }\href {\doibase
  https://doi.org/10.1364/OPTICA.6.000296} {\bibfield  {journal} {\bibinfo
  {journal} {Optica}\ }\textbf {\bibinfo {volume} {6}},\ \bibinfo {pages} {296}
  (\bibinfo {year} {2019})}\BibitemShut {NoStop}%
\bibitem [{\citenamefont {Tandaechanurat}\ \emph {et~al.}(2011)\citenamefont
  {Tandaechanurat}, \citenamefont {Ishida}, \citenamefont {Guimard},
  \citenamefont {Nomura}, \citenamefont {Iwamoto},\ and\ \citenamefont
  {Arakawa}}]{Tandaechanurat2011NP}%
  \BibitemOpen
  \bibfield  {author} {\bibinfo {author} {\bibfnamefont {A.}~\bibnamefont
  {Tandaechanurat}}, \bibinfo {author} {\bibfnamefont {S.}~\bibnamefont
  {Ishida}}, \bibinfo {author} {\bibfnamefont {D.}~\bibnamefont {Guimard}},
  \bibinfo {author} {\bibfnamefont {M.}~\bibnamefont {Nomura}}, \bibinfo
  {author} {\bibfnamefont {S.}~\bibnamefont {Iwamoto}}, \ and\ \bibinfo
  {author} {\bibfnamefont {Y.}~\bibnamefont {Arakawa}},\ }\href {\doibase
  10.1038/nphoton.2010.286} {\bibfield  {journal} {\bibinfo  {journal} {Nature
  Photonics}\ }\textbf {\bibinfo {volume} {5}},\ \bibinfo {pages} {91}
  (\bibinfo {year} {2011})}\BibitemShut {NoStop}%
\bibitem [{\citenamefont {Clerk}\ \emph {et~al.}(2010)\citenamefont {Clerk},
  \citenamefont {Devoret}, \citenamefont {Girvin}, \citenamefont {Marquardt},\
  and\ \citenamefont {Schoelkopf}}]{Clerk2010RMP}%
  \BibitemOpen
  \bibfield  {author} {\bibinfo {author} {\bibfnamefont {A.~A.}\ \bibnamefont
  {Clerk}}, \bibinfo {author} {\bibfnamefont {M.~H.}\ \bibnamefont {Devoret}},
  \bibinfo {author} {\bibfnamefont {S.~M.}\ \bibnamefont {Girvin}}, \bibinfo
  {author} {\bibfnamefont {F.}~\bibnamefont {Marquardt}}, \ and\ \bibinfo
  {author} {\bibfnamefont {R.~J.}\ \bibnamefont {Schoelkopf}},\ }\href@noop {}
  {\bibfield  {journal} {\bibinfo  {journal} {Rev. Mod. Phys.}\ }\textbf
  {\bibinfo {volume} {82}},\ \bibinfo {pages} {1155} (\bibinfo {year}
  {2010})}\BibitemShut {NoStop}%
\bibitem [{\citenamefont {Devashish}\ \emph {et~al.}(2017)\citenamefont
  {Devashish}, \citenamefont {Hasan}, \citenamefont {van~der Vegt},\ and\
  \citenamefont {Vos}}]{Devashish2017PRB}%
  \BibitemOpen
  \bibfield  {author} {\bibinfo {author} {\bibfnamefont {D.}~\bibnamefont
  {Devashish}}, \bibinfo {author} {\bibfnamefont {S.~B.}\ \bibnamefont
  {Hasan}}, \bibinfo {author} {\bibfnamefont {J.~J.}\ \bibnamefont {van~der
  Vegt}}, \ and\ \bibinfo {author} {\bibfnamefont {W.~L.}\ \bibnamefont
  {Vos}},\ }\href {\doibase 10.1103/PhysRevB.95.155141} {\bibfield  {journal}
  {\bibinfo  {journal} {Physical Review B}\ }\textbf {\bibinfo {volume} {95}},\
  \bibinfo {pages} {155141} (\bibinfo {year} {2017})}\BibitemShut {NoStop}%
\bibitem [{\citenamefont {L{\`{o}}pez}(2003)}]{Lopez2003AM}%
  \BibitemOpen
  \bibfield  {author} {\bibinfo {author} {\bibfnamefont {C.}~\bibnamefont
  {L{\`{o}}pez}},\ }\href {\doibase 101002/adma.200300386} {\bibfield
  {journal} {\bibinfo  {journal} {Advanced Materials}\ }\textbf {\bibinfo
  {volume} {15}},\ \bibinfo {pages} {1679} (\bibinfo {year}
  {2003})}\BibitemShut {NoStop}%
\bibitem [{\citenamefont {Benisty}\ and\ \citenamefont
  {Weisbuch}(2006)}]{Benisty2006ProgrOptics}%
  \BibitemOpen
  \bibfield  {author} {\bibinfo {author} {\bibfnamefont {H.}~\bibnamefont
  {Benisty}}\ and\ \bibinfo {author} {\bibfnamefont {C.}~\bibnamefont
  {Weisbuch}},\ }in\ \href@noop {} {\emph {\bibinfo {booktitle} {Progress in
  Optics 49}}},\ \bibinfo {editor} {edited by\ \bibinfo {editor} {\bibfnamefont
  {E.}~\bibnamefont {Wolf}}}\ (\bibinfo  {publisher} {Elsevier, Amsterdam},\
  \bibinfo {year} {2006})\ pp.\ \bibinfo {pages} {177--313}\BibitemShut
  {NoStop}%
\bibitem [{\citenamefont {Galisteo~L{\`{o}}pez}\ \emph
  {et~al.}(2011)\citenamefont {Galisteo~L{\`{o}}pez}, \citenamefont {Ibisate},
  \citenamefont {Sapienza}, \citenamefont {Froufe-P{\'{e}}rez}, \citenamefont
  {Blanco},\ and\ \citenamefont {L{\`{o}}pez}}]{Galisteo2011AM}%
  \BibitemOpen
  \bibfield  {author} {\bibinfo {author} {\bibfnamefont {J.~F.}\ \bibnamefont
  {Galisteo~L{\`{o}}pez}}, \bibinfo {author} {\bibfnamefont {M.}~\bibnamefont
  {Ibisate}}, \bibinfo {author} {\bibfnamefont {R.}~\bibnamefont {Sapienza}},
  \bibinfo {author} {\bibfnamefont {L.~S.}\ \bibnamefont {Froufe-P{\'{e}}rez}},
  \bibinfo {author} {\bibfnamefont {{\`{A}}.}~\bibnamefont {Blanco}}, \ and\
  \bibinfo {author} {\bibfnamefont {C.}~\bibnamefont {L{\`{o}}pez}},\
  }\href@noop {} {\bibfield  {journal} {\bibinfo  {journal} {Adv. Mater.}\
  }\textbf {\bibinfo {volume} {23}},\ \bibinfo {pages} {30} (\bibinfo {year}
  {2011})}\BibitemShut {NoStop}%
\bibitem [{\citenamefont {Ogawa}\ \emph {et~al.}(2004)\citenamefont {Ogawa},
  \citenamefont {Imada}, \citenamefont {Yoshimoto}, \citenamefont {Okano},\
  and\ \citenamefont {Noda}}]{Ogawa2004S}%
  \BibitemOpen
  \bibfield  {author} {\bibinfo {author} {\bibfnamefont {S.}~\bibnamefont
  {Ogawa}}, \bibinfo {author} {\bibfnamefont {M.}~\bibnamefont {Imada}},
  \bibinfo {author} {\bibfnamefont {S.}~\bibnamefont {Yoshimoto}}, \bibinfo
  {author} {\bibfnamefont {M.}~\bibnamefont {Okano}}, \ and\ \bibinfo {author}
  {\bibfnamefont {S.}~\bibnamefont {Noda}},\ }\href@noop {} {\bibfield
  {journal} {\bibinfo  {journal} {Science}\ }\textbf {\bibinfo {volume}
  {305}},\ \bibinfo {pages} {227} (\bibinfo {year} {2004})}\BibitemShut
  {NoStop}%
\bibitem [{\citenamefont {Lodahl}\ \emph {et~al.}(2004)\citenamefont {Lodahl},
  \citenamefont {Van~Driel}, \citenamefont {Nikolaev}, \citenamefont {Irman},
  \citenamefont {Overgaag}, \citenamefont {Vanmaekelbergh},\ and\ \citenamefont
  {Vos}}]{Lodahl2004Nat}%
  \BibitemOpen
  \bibfield  {author} {\bibinfo {author} {\bibfnamefont {P.}~\bibnamefont
  {Lodahl}}, \bibinfo {author} {\bibfnamefont {A.~F.}\ \bibnamefont
  {Van~Driel}}, \bibinfo {author} {\bibfnamefont {I.~S.}\ \bibnamefont
  {Nikolaev}}, \bibinfo {author} {\bibfnamefont {A.}~\bibnamefont {Irman}},
  \bibinfo {author} {\bibfnamefont {K.}~\bibnamefont {Overgaag}}, \bibinfo
  {author} {\bibfnamefont {D.}~\bibnamefont {Vanmaekelbergh}}, \ and\ \bibinfo
  {author} {\bibfnamefont {W.~L.}\ \bibnamefont {Vos}},\ }\href {\doibase
  10.1038/nature02772} {\bibfield  {journal} {\bibinfo  {journal} {Nature}\
  }\textbf {\bibinfo {volume} {430}},\ \bibinfo {pages} {654} (\bibinfo {year}
  {2004})}\BibitemShut {NoStop}%
\bibitem [{\citenamefont {Aoki}\ \emph {et~al.}(2008)\citenamefont {Aoki},
  \citenamefont {Guimard}, \citenamefont {Nishioka}, \citenamefont {Nomura},
  \citenamefont {Iwamoto},\ and\ \citenamefont {Arakawa}}]{Aoki2008NP}%
  \BibitemOpen
  \bibfield  {author} {\bibinfo {author} {\bibfnamefont {K.}~\bibnamefont
  {Aoki}}, \bibinfo {author} {\bibfnamefont {D.}~\bibnamefont {Guimard}},
  \bibinfo {author} {\bibfnamefont {M.}~\bibnamefont {Nishioka}}, \bibinfo
  {author} {\bibfnamefont {M.}~\bibnamefont {Nomura}}, \bibinfo {author}
  {\bibfnamefont {S.}~\bibnamefont {Iwamoto}}, \ and\ \bibinfo {author}
  {\bibfnamefont {Y.}~\bibnamefont {Arakawa}},\ }\href {\doibase
  10.1038/nphoton.2008.202} {\bibfield  {journal} {\bibinfo  {journal} {Nature
  Photonics}\ }\textbf {\bibinfo {volume} {2}},\ \bibinfo {pages} {688}
  (\bibinfo {year} {2008})}\BibitemShut {NoStop}%
\bibitem [{\citenamefont {Leistikow}\ \emph {et~al.}(2011)\citenamefont
  {Leistikow}, \citenamefont {Mosk}, \citenamefont {Yeganegi}, \citenamefont
  {Huisman},\ and\ \citenamefont {Vos}}]{Leistikow2011PRL}%
  \BibitemOpen
  \bibfield  {author} {\bibinfo {author} {\bibfnamefont {M.~D.}\ \bibnamefont
  {Leistikow}}, \bibinfo {author} {\bibfnamefont {A.~P.}\ \bibnamefont {Mosk}},
  \bibinfo {author} {\bibfnamefont {E.}~\bibnamefont {Yeganegi}}, \bibinfo
  {author} {\bibfnamefont {A.}~\bibnamefont {Huisman}, \bibfnamefont
  {S~R~Lagendijk}}, \ and\ \bibinfo {author} {\bibfnamefont {W.~L.}\
  \bibnamefont {Vos}},\ }\href@noop {} {\bibfield  {journal} {\bibinfo
  {journal} {Physical Review Letters}\ }\textbf {\bibinfo {volume} {107}},\
  \bibinfo {pages} {193903} (\bibinfo {year} {2011})}\BibitemShut {NoStop}%
\bibitem [{\citenamefont {Koenderink}\ \emph {et~al.}(2002)\citenamefont
  {Koenderink}, \citenamefont {Bechger}, \citenamefont {Schriemer},
  \citenamefont {Lagendijk},\ and\ \citenamefont {Vos}}]{Koenderink2002PRL}%
  \BibitemOpen
  \bibfield  {author} {\bibinfo {author} {\bibfnamefont {A.~F.}\ \bibnamefont
  {Koenderink}}, \bibinfo {author} {\bibfnamefont {L.}~\bibnamefont {Bechger}},
  \bibinfo {author} {\bibfnamefont {H.}~\bibnamefont {Schriemer}}, \bibinfo
  {author} {\bibfnamefont {A.}~\bibnamefont {Lagendijk}}, \ and\ \bibinfo
  {author} {\bibfnamefont {W.~L.}\ \bibnamefont {Vos}},\ }\href@noop {}
  {\bibfield  {journal} {\bibinfo  {journal} {Phys. Rev. Lett.}\ }\textbf
  {\bibinfo {volume} {88}},\ \bibinfo {pages} {143903} (\bibinfo {year}
  {2002})}\BibitemShut {NoStop}%
\bibitem [{\citenamefont {Koenderink}\ \emph {et~al.}(2003)\citenamefont
  {Koenderink}, \citenamefont {Bechger}, \citenamefont {Lagendijk},\ and\
  \citenamefont {Vos}}]{Koenderink2003PSSA}%
  \BibitemOpen
  \bibfield  {author} {\bibinfo {author} {\bibfnamefont {A.~F.}\ \bibnamefont
  {Koenderink}}, \bibinfo {author} {\bibfnamefont {L.}~\bibnamefont {Bechger}},
  \bibinfo {author} {\bibfnamefont {A.}~\bibnamefont {Lagendijk}}, \ and\
  \bibinfo {author} {\bibfnamefont {W.~L.}\ \bibnamefont {Vos}},\ }\href
  {\doibase 10.1002/pssa.200303115} {\bibfield  {journal} {\bibinfo  {journal}
  {Phys. Stat. Sol. (a)}\ }\textbf {\bibinfo {volume} {197}},\ \bibinfo {pages}
  {648} (\bibinfo {year} {2003})}\BibitemShut {NoStop}%
\bibitem [{\citenamefont {Hasan}\ \emph {et~al.}(2018)\citenamefont {Hasan},
  \citenamefont {Mosk}, \citenamefont {Vos},\ and\ \citenamefont
  {Lagendijk}}]{Hasan2018PRL}%
  \BibitemOpen
  \bibfield  {author} {\bibinfo {author} {\bibfnamefont {S.~B.}\ \bibnamefont
  {Hasan}}, \bibinfo {author} {\bibfnamefont {A.~P.}\ \bibnamefont {Mosk}},
  \bibinfo {author} {\bibfnamefont {W.~L.}\ \bibnamefont {Vos}}, \ and\
  \bibinfo {author} {\bibfnamefont {A.}~\bibnamefont {Lagendijk}},\ }\href@noop
  {} {\bibfield  {journal} {\bibinfo  {journal} {Phys. Rev. Lett.}\ }\textbf
  {\bibinfo {volume} {120}},\ \bibinfo {pages} {237402} (\bibinfo {year}
  {2018})}\BibitemShut {NoStop}%
\bibitem [{\citenamefont {Lin}\ \emph {et~al.}(1998)\citenamefont {Lin},
  \citenamefont {Fleming}, \citenamefont {Hetherington}, \citenamefont {Smith},
  \citenamefont {Biswas}, \citenamefont {Ho}, \citenamefont {Sigalas},
  \citenamefont {Zubrzycki}, \citenamefont {Kurtz},\ and\ \citenamefont
  {Bur}}]{Lin1998Nature}%
  \BibitemOpen
  \bibfield  {author} {\bibinfo {author} {\bibfnamefont {S.~Y.}\ \bibnamefont
  {Lin}}, \bibinfo {author} {\bibfnamefont {J.~G.}\ \bibnamefont {Fleming}},
  \bibinfo {author} {\bibfnamefont {D.~L.}\ \bibnamefont {Hetherington}},
  \bibinfo {author} {\bibfnamefont {B.~K.}\ \bibnamefont {Smith}}, \bibinfo
  {author} {\bibfnamefont {R.}~\bibnamefont {Biswas}}, \bibinfo {author}
  {\bibfnamefont {K.~M.}\ \bibnamefont {Ho}}, \bibinfo {author} {\bibfnamefont
  {M.~M.}\ \bibnamefont {Sigalas}}, \bibinfo {author} {\bibfnamefont
  {W.}~\bibnamefont {Zubrzycki}}, \bibinfo {author} {\bibfnamefont {S.~R.}\
  \bibnamefont {Kurtz}}, \ and\ \bibinfo {author} {\bibfnamefont
  {J.}~\bibnamefont {Bur}},\ }\href {\doibase 10.1038/28343} {\bibfield
  {journal} {\bibinfo  {journal} {Nature}\ }\textbf {\bibinfo {volume} {394}},\
  \bibinfo {pages} {251} (\bibinfo {year} {1998})}\BibitemShut {NoStop}%
\bibitem [{\citenamefont {Thijssen}\ \emph {et~al.}(1999)\citenamefont
  {Thijssen}, \citenamefont {Sprik}, \citenamefont {Wijnhoven}, \citenamefont
  {Megens}, \citenamefont {Narayanan}, \citenamefont {Lagendijk},\ and\
  \citenamefont {Vos}}]{Thijssen1999PRL}%
  \BibitemOpen
  \bibfield  {author} {\bibinfo {author} {\bibfnamefont {M.~S.}\ \bibnamefont
  {Thijssen}}, \bibinfo {author} {\bibfnamefont {R.}~\bibnamefont {Sprik}},
  \bibinfo {author} {\bibfnamefont {J.~E. G.~J.}\ \bibnamefont {Wijnhoven}},
  \bibinfo {author} {\bibfnamefont {M.}~\bibnamefont {Megens}}, \bibinfo
  {author} {\bibfnamefont {T.}~\bibnamefont {Narayanan}}, \bibinfo {author}
  {\bibfnamefont {A.}~\bibnamefont {Lagendijk}}, \ and\ \bibinfo {author}
  {\bibfnamefont {W.~L.}\ \bibnamefont {Vos}},\ }\href@noop {} {\bibfield
  {journal} {\bibinfo  {journal} {Phys. Rev. Lett.}\ }\textbf {\bibinfo
  {volume} {83}},\ \bibinfo {pages} {2730} (\bibinfo {year}
  {1999})}\BibitemShut {NoStop}%
\bibitem [{\citenamefont {Noda}\ \emph {et~al.}(2000)\citenamefont {Noda},
  \citenamefont {Tomoda}, \citenamefont {Yamamoto},\ and\ \citenamefont
  {Chutinan}}]{Noda2000Science}%
  \BibitemOpen
  \bibfield  {author} {\bibinfo {author} {\bibfnamefont {S.}~\bibnamefont
  {Noda}}, \bibinfo {author} {\bibfnamefont {K.}~\bibnamefont {Tomoda}},
  \bibinfo {author} {\bibfnamefont {N.}~\bibnamefont {Yamamoto}}, \ and\
  \bibinfo {author} {\bibfnamefont {A.}~\bibnamefont {Chutinan}},\ }\href
  {\doibase 10.1126/science.289.5479.604} {\bibfield  {journal} {\bibinfo
  {journal} {Science}\ }\textbf {\bibinfo {volume} {289}},\ \bibinfo {pages}
  {604} (\bibinfo {year} {2000})}\BibitemShut {NoStop}%
\bibitem [{\citenamefont {Blanco}\ \emph {et~al.}(2000)\citenamefont {Blanco},
  \citenamefont {Chomski}, \citenamefont {Grabtchak}, \citenamefont {Ibisate},
  \citenamefont {John}, \citenamefont {Leonard}, \citenamefont {L{\`{o}}pez},
  \citenamefont {Meseguer}, \citenamefont {Miguez}, \citenamefont {Mondla},
  \citenamefont {Ozin}, \citenamefont {Toader},\ and\ \citenamefont {van
  Driel}}]{Blanco2000Nature}%
  \BibitemOpen
  \bibfield  {author} {\bibinfo {author} {\bibfnamefont {A.}~\bibnamefont
  {Blanco}}, \bibinfo {author} {\bibfnamefont {E.}~\bibnamefont {Chomski}},
  \bibinfo {author} {\bibfnamefont {S.}~\bibnamefont {Grabtchak}}, \bibinfo
  {author} {\bibfnamefont {M.}~\bibnamefont {Ibisate}}, \bibinfo {author}
  {\bibfnamefont {S.}~\bibnamefont {John}}, \bibinfo {author} {\bibfnamefont
  {S.~W.}\ \bibnamefont {Leonard}}, \bibinfo {author} {\bibfnamefont
  {C.}~\bibnamefont {L{\`{o}}pez}}, \bibinfo {author} {\bibfnamefont
  {F.}~\bibnamefont {Meseguer}}, \bibinfo {author} {\bibfnamefont
  {H.}~\bibnamefont {Miguez}}, \bibinfo {author} {\bibfnamefont {J.~P.}\
  \bibnamefont {Mondla}}, \bibinfo {author} {\bibfnamefont {G.~A.}\
  \bibnamefont {Ozin}}, \bibinfo {author} {\bibfnamefont {O.}~\bibnamefont
  {Toader}}, \ and\ \bibinfo {author} {\bibfnamefont {H.~M.}\ \bibnamefont {van
  Driel}},\ }\href {\doibase 10.1038/35013024} {\bibfield  {journal} {\bibinfo
  {journal} {Nature}\ }\textbf {\bibinfo {volume} {405}},\ \bibinfo {pages}
  {437} (\bibinfo {year} {2000})}\BibitemShut {NoStop}%
\bibitem [{\citenamefont {Vlasov}\ \emph {et~al.}(2001)\citenamefont {Vlasov},
  \citenamefont {Bo}, \citenamefont {Sturm},\ and\ \citenamefont
  {Norris}}]{Vlasov2001N}%
  \BibitemOpen
  \bibfield  {author} {\bibinfo {author} {\bibfnamefont {Y.~A.}\ \bibnamefont
  {Vlasov}}, \bibinfo {author} {\bibfnamefont {X.-Z.}\ \bibnamefont {Bo}},
  \bibinfo {author} {\bibfnamefont {J.~C.}\ \bibnamefont {Sturm}}, \ and\
  \bibinfo {author} {\bibfnamefont {D.~J.}\ \bibnamefont {Norris}},\
  }\href@noop {} {\bibfield  {journal} {\bibinfo  {journal} {Nature}\ }\textbf
  {\bibinfo {volume} {414}},\ \bibinfo {pages} {289} (\bibinfo {year}
  {2001})}\BibitemShut {NoStop}%
\bibitem [{\citenamefont {Schilling}\ \emph {et~al.}(2005)\citenamefont
  {Schilling}, \citenamefont {White}, \citenamefont {Scherer}, \citenamefont
  {Stupian}, \citenamefont {Hillebrand},\ and\ \citenamefont
  {G{\"{o}}sele}}]{Schilling2005APL}%
  \BibitemOpen
  \bibfield  {author} {\bibinfo {author} {\bibfnamefont {J.}~\bibnamefont
  {Schilling}}, \bibinfo {author} {\bibfnamefont {J.}~\bibnamefont {White}},
  \bibinfo {author} {\bibfnamefont {A.}~\bibnamefont {Scherer}}, \bibinfo
  {author} {\bibfnamefont {G.}~\bibnamefont {Stupian}}, \bibinfo {author}
  {\bibfnamefont {R.}~\bibnamefont {Hillebrand}}, \ and\ \bibinfo {author}
  {\bibfnamefont {U.}~\bibnamefont {G{\"{o}}sele}},\ }\href {\doibase
  10.1063/1.1842855} {\bibfield  {journal} {\bibinfo  {journal} {Applied
  Physics Letters}\ }\textbf {\bibinfo {volume} {86}},\ \bibinfo {pages}
  {011101} (\bibinfo {year} {2005})}\BibitemShut {NoStop}%
\bibitem [{\citenamefont {Garc{\'{i}}a-Santamar{\'{i}}a}\ \emph
  {et~al.}(2007)\citenamefont {Garc{\'{i}}a-Santamar{\'{i}}a}, \citenamefont
  {Xu}, \citenamefont {Lousse}, \citenamefont {Fan}, \citenamefont {Braun},\
  and\ \citenamefont {Lewis}}]{Garcia-Santamaria2007AM}%
  \BibitemOpen
  \bibfield  {author} {\bibinfo {author} {\bibfnamefont {F.}~\bibnamefont
  {Garc{\'{i}}a-Santamar{\'{i}}a}}, \bibinfo {author} {\bibfnamefont
  {M.}~\bibnamefont {Xu}}, \bibinfo {author} {\bibfnamefont {V.}~\bibnamefont
  {Lousse}}, \bibinfo {author} {\bibfnamefont {S.}~\bibnamefont {Fan}},
  \bibinfo {author} {\bibfnamefont {P.~V.}\ \bibnamefont {Braun}}, \ and\
  \bibinfo {author} {\bibfnamefont {J.~A.}\ \bibnamefont {Lewis}},\ }\href
  {\doibase 10.1002/adma.200602906} {\bibfield  {journal} {\bibinfo  {journal}
  {Advanced Materials}\ }\textbf {\bibinfo {volume} {19}},\ \bibinfo {pages}
  {1567} (\bibinfo {year} {2007})}\BibitemShut {NoStop}%
\bibitem [{\citenamefont {Takahashi}\ \emph {et~al.}(2009)\citenamefont
  {Takahashi}, \citenamefont {Suzuki}, \citenamefont {Okano}, \citenamefont
  {Imada}, \citenamefont {Nakamori}, \citenamefont {Ota}, \citenamefont
  {Ishizaki},\ and\ \citenamefont {Noda}}]{Takahashi2009NM}%
  \BibitemOpen
  \bibfield  {author} {\bibinfo {author} {\bibfnamefont {S.}~\bibnamefont
  {Takahashi}}, \bibinfo {author} {\bibfnamefont {K.}~\bibnamefont {Suzuki}},
  \bibinfo {author} {\bibfnamefont {M.}~\bibnamefont {Okano}}, \bibinfo
  {author} {\bibfnamefont {M.}~\bibnamefont {Imada}}, \bibinfo {author}
  {\bibfnamefont {T.}~\bibnamefont {Nakamori}}, \bibinfo {author}
  {\bibfnamefont {Y.}~\bibnamefont {Ota}}, \bibinfo {author} {\bibfnamefont
  {K.}~\bibnamefont {Ishizaki}}, \ and\ \bibinfo {author} {\bibfnamefont
  {S.}~\bibnamefont {Noda}},\ }\href {\doibase 10.1038/nmat2507} {\bibfield
  {journal} {\bibinfo  {journal} {Nature Materials}\ }\textbf {\bibinfo
  {volume} {8}},\ \bibinfo {pages} {721} (\bibinfo {year} {2009})}\BibitemShut
  {NoStop}%
\bibitem [{\citenamefont {Staude}\ \emph {et~al.}(2010)\citenamefont {Staude},
  \citenamefont {Thiel}, \citenamefont {Essig}, \citenamefont {Wolff},
  \citenamefont {Busch}, \citenamefont {von Freymann},\ and\ \citenamefont
  {Wegener}}]{Staude2010OL}%
  \BibitemOpen
  \bibfield  {author} {\bibinfo {author} {\bibfnamefont {I.}~\bibnamefont
  {Staude}}, \bibinfo {author} {\bibfnamefont {M.}~\bibnamefont {Thiel}},
  \bibinfo {author} {\bibfnamefont {S.}~\bibnamefont {Essig}}, \bibinfo
  {author} {\bibfnamefont {C.}~\bibnamefont {Wolff}}, \bibinfo {author}
  {\bibfnamefont {K.}~\bibnamefont {Busch}}, \bibinfo {author} {\bibfnamefont
  {G.}~\bibnamefont {von Freymann}}, \ and\ \bibinfo {author} {\bibfnamefont
  {M.}~\bibnamefont {Wegener}},\ }\href {\doibase 10.1364/OL.35.001094}
  {\bibfield  {journal} {\bibinfo  {journal} {Opt. Lett.}\ }\textbf {\bibinfo
  {volume} {35}},\ \bibinfo {pages} {1094} (\bibinfo {year}
  {2010})}\BibitemShut {NoStop}%
\bibitem [{\citenamefont {Huisman}\ \emph {et~al.}(2011)\citenamefont
  {Huisman}, \citenamefont {Nair}, \citenamefont {Woldering}, \citenamefont
  {Leistikow}, \citenamefont {Mosk},\ and\ \citenamefont
  {Vos}}]{Huisman2011PRB}%
  \BibitemOpen
  \bibfield  {author} {\bibinfo {author} {\bibfnamefont {S.~R.}\ \bibnamefont
  {Huisman}}, \bibinfo {author} {\bibfnamefont {R.~V.}\ \bibnamefont {Nair}},
  \bibinfo {author} {\bibfnamefont {L.~A.}\ \bibnamefont {Woldering}}, \bibinfo
  {author} {\bibfnamefont {M.~D.}\ \bibnamefont {Leistikow}}, \bibinfo {author}
  {\bibfnamefont {A.~P.}\ \bibnamefont {Mosk}}, \ and\ \bibinfo {author}
  {\bibfnamefont {W.~L.}\ \bibnamefont {Vos}},\ }\href {\doibase
  10.1103/PhysRevB.83.205313} {\bibfield  {journal} {\bibinfo  {journal}
  {Physical Review B}\ }\textbf {\bibinfo {volume} {83}},\ \bibinfo {pages}
  {205313} (\bibinfo {year} {2011})}\BibitemShut {NoStop}%
\bibitem [{\citenamefont {Fr{\"{o}}lich}\ \emph {et~al.}(2013)\citenamefont
  {Fr{\"{o}}lich}, \citenamefont {Fischer}, \citenamefont {Zebrowski},
  \citenamefont {Busch},\ and\ \citenamefont {Wegener}}]{Frolich2013AM}%
  \BibitemOpen
  \bibfield  {author} {\bibinfo {author} {\bibfnamefont {A.}~\bibnamefont
  {Fr{\"{o}}lich}}, \bibinfo {author} {\bibfnamefont {J.}~\bibnamefont
  {Fischer}}, \bibinfo {author} {\bibfnamefont {T.}~\bibnamefont {Zebrowski}},
  \bibinfo {author} {\bibfnamefont {K.}~\bibnamefont {Busch}}, \ and\ \bibinfo
  {author} {\bibfnamefont {M.}~\bibnamefont {Wegener}},\ }\href {\doibase
  10.1002/adma.201300896} {\bibfield  {journal} {\bibinfo  {journal} {Advanced
  Materials}\ }\textbf {\bibinfo {volume} {25}},\ \bibinfo {pages} {3588 }
  (\bibinfo {year} {2013})}\BibitemShut {NoStop}%
\bibitem [{\citenamefont {Marichy}\ \emph {et~al.}(2016)\citenamefont
  {Marichy}, \citenamefont {Muller}, \citenamefont {Froufe-P{\'{e}}rez},\ and\
  \citenamefont {Scheffold}}]{Marichy2016SR}%
  \BibitemOpen
  \bibfield  {author} {\bibinfo {author} {\bibfnamefont {C.}~\bibnamefont
  {Marichy}}, \bibinfo {author} {\bibfnamefont {N.}~\bibnamefont {Muller}},
  \bibinfo {author} {\bibfnamefont {L.~S.}\ \bibnamefont {Froufe-P{\'{e}}rez}},
  \ and\ \bibinfo {author} {\bibfnamefont {F.}~\bibnamefont {Scheffold}},\
  }\href {\doibase 10.1038/srep21818} {\bibfield  {journal} {\bibinfo
  {journal} {Scientific Reports}\ }\textbf {\bibinfo {volume} {6}},\ \bibinfo
  {pages} {1} (\bibinfo {year} {2016})}\BibitemShut {NoStop}%
\bibitem [{\citenamefont {Robertson}\ \emph {et~al.}(1992)\citenamefont
  {Robertson}, \citenamefont {Arjavalingam}, \citenamefont {Meade},
  \citenamefont {Brommer}, \citenamefont {Rappe},\ and\ \citenamefont
  {Joannopoulos}}]{Robertson1992PRL}%
  \BibitemOpen
  \bibfield  {author} {\bibinfo {author} {\bibfnamefont {W.~M.}\ \bibnamefont
  {Robertson}}, \bibinfo {author} {\bibfnamefont {G.}~\bibnamefont
  {Arjavalingam}}, \bibinfo {author} {\bibfnamefont {R.~D.}\ \bibnamefont
  {Meade}}, \bibinfo {author} {\bibfnamefont {K.~D.}\ \bibnamefont {Brommer}},
  \bibinfo {author} {\bibfnamefont {A.~M.}\ \bibnamefont {Rappe}}, \ and\
  \bibinfo {author} {\bibfnamefont {J.~D.}\ \bibnamefont {Joannopoulos}},\
  }\href@noop {} {\bibfield  {journal} {\bibinfo  {journal} {Phys. Rev. Lett.}\
  }\textbf {\bibinfo {volume} {68}},\ \bibinfo {pages} {2023} (\bibinfo {year}
  {1992})}\BibitemShut {NoStop}%
\bibitem [{\citenamefont {Sakoda}(2005)}]{Sakoda2005book}%
  \BibitemOpen
  \bibfield  {author} {\bibinfo {author} {\bibfnamefont {K.}~\bibnamefont
  {Sakoda}},\ }\href@noop {} {\emph {\bibinfo {title} {Optical properties of
  photonic crystals}}}\ (\bibinfo  {publisher} {Springer Verlag,
  Heidelberg-Berlin},\ \bibinfo {year} {2005})\BibitemShut {NoStop}%
\bibitem [{\citenamefont {Li}\ and\ \citenamefont {Zhang}(2002)}]{Li2000PRB}%
  \BibitemOpen
  \bibfield  {author} {\bibinfo {author} {\bibfnamefont {Z.-Y.}\ \bibnamefont
  {Li}}\ and\ \bibinfo {author} {\bibnamefont {Zhang}},\ }\href@noop {}
  {\bibfield  {journal} {\bibinfo  {journal} {Phys. Rev. B}\ }\textbf {\bibinfo
  {volume} {62}},\ \bibinfo {pages} {1516} (\bibinfo {year}
  {2002})}\BibitemShut {NoStop}%
\bibitem [{\citenamefont {Wang}\ \emph {et~al.}(2003)\citenamefont {Wang},
  \citenamefont {Chan}, \citenamefont {Zhang}, \citenamefont {Chen},
  \citenamefont {Ming},\ and\ \citenamefont {Sheng}}]{Wang2003PRE}%
  \BibitemOpen
  \bibfield  {author} {\bibinfo {author} {\bibfnamefont {Z.~L.}\ \bibnamefont
  {Wang}}, \bibinfo {author} {\bibfnamefont {C.~T.}\ \bibnamefont {Chan}},
  \bibinfo {author} {\bibfnamefont {W.~Y.}\ \bibnamefont {Zhang}}, \bibinfo
  {author} {\bibfnamefont {Z.}~\bibnamefont {Chen}}, \bibinfo {author}
  {\bibfnamefont {N.~B.}\ \bibnamefont {Ming}}, \ and\ \bibinfo {author}
  {\bibfnamefont {P.}~\bibnamefont {Sheng}},\ }\href@noop {} {\bibfield
  {journal} {\bibinfo  {journal} {Phys. Rev. E}\ }\textbf {\bibinfo {volume}
  {67}},\ \bibinfo {pages} {016612} (\bibinfo {year} {2003})}\BibitemShut
  {NoStop}%
\bibitem [{\citenamefont {Grishina}\ \emph {et~al.}(2018)\citenamefont
  {Grishina}, \citenamefont {Harteveld}, \citenamefont {Pacureanu},
  \citenamefont {Devashish}, \citenamefont {Lagendijk}, \citenamefont
  {Cloetens},\ and\ \citenamefont {Vos}}]{Grishina2018Arxiv}%
  \BibitemOpen
  \bibfield  {author} {\bibinfo {author} {\bibfnamefont {D.~A.}\ \bibnamefont
  {Grishina}}, \bibinfo {author} {\bibfnamefont {C.~A.~M.}\ \bibnamefont
  {Harteveld}}, \bibinfo {author} {\bibfnamefont {A.}~\bibnamefont
  {Pacureanu}}, \bibinfo {author} {\bibfnamefont {D.}~\bibnamefont
  {Devashish}}, \bibinfo {author} {\bibfnamefont {A.}~\bibnamefont
  {Lagendijk}}, \bibinfo {author} {\bibfnamefont {P.}~\bibnamefont {Cloetens}},
  \ and\ \bibinfo {author} {\bibfnamefont {W.~L.}\ \bibnamefont {Vos}},\
  }\href@noop {} {\bibfield  {journal} {\bibinfo  {journal} {arXiv:1808.01392}\
  } (\bibinfo {year} {2018})},\ \Eprint {http://arxiv.org/abs/1808.01392}
  {arXiv:1808.01392} \BibitemShut {NoStop}%
\bibitem [{\citenamefont {Ho}\ \emph {et~al.}(1994)\citenamefont {Ho},
  \citenamefont {Chan}, \citenamefont {Soukoulis}, \citenamefont {Biswas},\
  and\ \citenamefont {Sigalas}}]{Ho1994SSC}%
  \BibitemOpen
  \bibfield  {author} {\bibinfo {author} {\bibfnamefont {K.~M.}\ \bibnamefont
  {Ho}}, \bibinfo {author} {\bibfnamefont {C.~T.}\ \bibnamefont {Chan}},
  \bibinfo {author} {\bibfnamefont {C.~M.}\ \bibnamefont {Soukoulis}}, \bibinfo
  {author} {\bibfnamefont {R.}~\bibnamefont {Biswas}}, \ and\ \bibinfo {author}
  {\bibfnamefont {M.}~\bibnamefont {Sigalas}},\ }\href {\doibase
  10.1016/0038-1098(94)90202-X} {\bibfield  {journal} {\bibinfo  {journal}
  {Solid State Communications}\ }\textbf {\bibinfo {volume} {89}},\ \bibinfo
  {pages} {413 } (\bibinfo {year} {1994})}\BibitemShut {NoStop}%
\bibitem [{\citenamefont {nanocops}(2012)}]{COPS2012youtube}%
  \BibitemOpen
  \bibfield  {author} {\bibinfo {author} {\bibnamefont {nanocops}},\ }\href
  {https://www.youtube.com/watch?v=MH0WGqHI1ss} {\enquote {\bibinfo {title}
  {{Animation: 3D Photonic Crystal with a Diamond Structure}},}\ } (\bibinfo
  {year} {2012})\BibitemShut {NoStop}%
\bibitem [{\citenamefont {Maldovan}\ and\ \citenamefont
  {Thomas}(2004)}]{Maldovan2004NM}%
  \BibitemOpen
  \bibfield  {author} {\bibinfo {author} {\bibfnamefont {M.}~\bibnamefont
  {Maldovan}}\ and\ \bibinfo {author} {\bibfnamefont {E.~L.}\ \bibnamefont
  {Thomas}},\ }\href {\doibase 10.1038/nmat1201} {\bibfield  {journal}
  {\bibinfo  {journal} {Nature Materials}\ }\textbf {\bibinfo {volume} {3}},\
  \bibinfo {pages} {593} (\bibinfo {year} {2004})}\BibitemShut {NoStop}%
\bibitem [{\citenamefont {Hillebrand}\ \emph {et~al.}(2003)\citenamefont
  {Hillebrand}, \citenamefont {Senz}, \citenamefont {Hergert},\ and\
  \citenamefont {G{\"{o}}sele}}]{Hillebrand2003JAP}%
  \BibitemOpen
  \bibfield  {author} {\bibinfo {author} {\bibfnamefont {R.}~\bibnamefont
  {Hillebrand}}, \bibinfo {author} {\bibfnamefont {S.}~\bibnamefont {Senz}},
  \bibinfo {author} {\bibfnamefont {W.}~\bibnamefont {Hergert}}, \ and\
  \bibinfo {author} {\bibfnamefont {U.}~\bibnamefont {G{\"{o}}sele}},\ }\href
  {\doibase 10.1063/1.1593796} {\bibfield  {journal} {\bibinfo  {journal}
  {Journal of Applied Physics}\ }\textbf {\bibinfo {volume} {94}},\ \bibinfo
  {pages} {2758} (\bibinfo {year} {2003})}\BibitemShut {NoStop}%
\bibitem [{\citenamefont {Woldering}\ \emph {et~al.}(2009)\citenamefont
  {Woldering}, \citenamefont {Mosk}, \citenamefont {Tjerkstra},\ and\
  \citenamefont {Vos}}]{Woldering2009JAP}%
  \BibitemOpen
  \bibfield  {author} {\bibinfo {author} {\bibfnamefont {L.~A.}\ \bibnamefont
  {Woldering}}, \bibinfo {author} {\bibfnamefont {A.~P.}\ \bibnamefont {Mosk}},
  \bibinfo {author} {\bibfnamefont {R.~W.}\ \bibnamefont {Tjerkstra}}, \ and\
  \bibinfo {author} {\bibfnamefont {W.~L.}\ \bibnamefont {Vos}},\ }\href
  {\doibase 10.1063/1.3103777} {\bibfield  {journal} {\bibinfo  {journal}
  {Journal of Applied Physics}\ }\textbf {\bibinfo {volume} {105}},\ \bibinfo
  {pages} {093108} (\bibinfo {year} {2009})}\BibitemShut {NoStop}%
\bibitem [{\citenamefont {Vos}\ \emph {et~al.}(1996)\citenamefont {Vos},
  \citenamefont {Sprik}, \citenamefont {van Blaaderen}, \citenamefont {Imhof},
  \citenamefont {Lagendijk},\ and\ \citenamefont {Wegdam}}]{Vos1996PRB}%
  \BibitemOpen
  \bibfield  {author} {\bibinfo {author} {\bibfnamefont {W.~L.}\ \bibnamefont
  {Vos}}, \bibinfo {author} {\bibfnamefont {R.}~\bibnamefont {Sprik}}, \bibinfo
  {author} {\bibfnamefont {A.}~\bibnamefont {van Blaaderen}}, \bibinfo {author}
  {\bibfnamefont {A.}~\bibnamefont {Imhof}}, \bibinfo {author} {\bibfnamefont
  {A.}~\bibnamefont {Lagendijk}}, \ and\ \bibinfo {author} {\bibfnamefont
  {G.~H.}\ \bibnamefont {Wegdam}},\ }\href@noop {} {\bibfield  {journal}
  {\bibinfo  {journal} {Phys. Rev. B}\ }\textbf {\bibinfo {volume} {53}},\
  \bibinfo {pages} {16231} (\bibinfo {year} {1996})}\BibitemShut {NoStop}%
\bibitem [{\citenamefont {Datta}\ \emph {et~al.}(1993)\citenamefont {Datta},
  \citenamefont {Chan}, \citenamefont {Ho},\ and\ \citenamefont
  {Soukoulis}}]{Datta1993PRB}%
  \BibitemOpen
  \bibfield  {author} {\bibinfo {author} {\bibfnamefont {S.}~\bibnamefont
  {Datta}}, \bibinfo {author} {\bibfnamefont {C.~T.}\ \bibnamefont {Chan}},
  \bibinfo {author} {\bibfnamefont {K.~M.}\ \bibnamefont {Ho}}, \ and\ \bibinfo
  {author} {\bibfnamefont {C.~M.}\ \bibnamefont {Soukoulis}},\ }\href@noop {}
  {\bibfield  {journal} {\bibinfo  {journal} {Physical Review B}\ }\textbf
  {\bibinfo {volume} {48}},\ \bibinfo {pages} {14936} (\bibinfo {year}
  {1993})}\BibitemShut {NoStop}%
\bibitem [{\citenamefont {van~den Broek}\ \emph {et~al.}(2012)\citenamefont
  {van~den Broek}, \citenamefont {Woldering}, \citenamefont {Tjerkstra},
  \citenamefont {Segerink}, \citenamefont {Setija},\ and\ \citenamefont
  {Vos}}]{VanDenBroek2012AFM}%
  \BibitemOpen
  \bibfield  {author} {\bibinfo {author} {\bibfnamefont {J.~M.}\ \bibnamefont
  {van~den Broek}}, \bibinfo {author} {\bibfnamefont {L.~A.}\ \bibnamefont
  {Woldering}}, \bibinfo {author} {\bibfnamefont {R.~W.}\ \bibnamefont
  {Tjerkstra}}, \bibinfo {author} {\bibfnamefont {F.~B.}\ \bibnamefont
  {Segerink}}, \bibinfo {author} {\bibfnamefont {I.~D.}\ \bibnamefont
  {Setija}}, \ and\ \bibinfo {author} {\bibfnamefont {W.~L.}\ \bibnamefont
  {Vos}},\ }\href {\doibase 10.1002/adfm.201101101} {\bibfield  {journal}
  {\bibinfo  {journal} {Advanced Functional Materials}\ }\textbf {\bibinfo
  {volume} {22}},\ \bibinfo {pages} {25} (\bibinfo {year} {2012})}\BibitemShut
  {NoStop}%
\bibitem [{\citenamefont {Grishina}(2017)}]{Grishina20173DNanophotonics}%
  \BibitemOpen
  \bibfield  {author} {\bibinfo {author} {\bibfnamefont {D.~A.}\ \bibnamefont
  {Grishina}},\ }\emph {\bibinfo {title} {{3D Silicon Nanophotonics}}},\ \href
  {https://doi.org/10.3990/1.9789036543743} {Ph.D. thesis},\ \bibinfo  {school}
  {University of Twente} (\bibinfo {year} {2017})\BibitemShut {NoStop}%
\bibitem [{\citenamefont {Tjerkstra}\ \emph {et~al.}(2011)\citenamefont
  {Tjerkstra}, \citenamefont {Woldering}, \citenamefont {van~den Broek},
  \citenamefont {Roozeboom}, \citenamefont {Setija},\ and\ \citenamefont
  {Vos}}]{Tjerkstra2011JVSTB}%
  \BibitemOpen
  \bibfield  {author} {\bibinfo {author} {\bibfnamefont {R.~W.}\ \bibnamefont
  {Tjerkstra}}, \bibinfo {author} {\bibfnamefont {L.~A.}\ \bibnamefont
  {Woldering}}, \bibinfo {author} {\bibfnamefont {J.~M.}\ \bibnamefont {van~den
  Broek}}, \bibinfo {author} {\bibfnamefont {F.}~\bibnamefont {Roozeboom}},
  \bibinfo {author} {\bibfnamefont {I.~D.}\ \bibnamefont {Setija}}, \ and\
  \bibinfo {author} {\bibfnamefont {W.~L.}\ \bibnamefont {Vos}},\ }\href@noop
  {} {\bibfield  {journal} {\bibinfo  {journal} {J. Vac. Sci. Technol. B}\
  }\textbf {\bibinfo {volume} {29}},\ \bibinfo {pages} {061604} (\bibinfo
  {year} {2011})}\BibitemShut {NoStop}%
\bibitem [{\citenamefont {Grishina}\ \emph {et~al.}(2015)\citenamefont
  {Grishina}, \citenamefont {Harteveld}, \citenamefont {Woldering},\ and\
  \citenamefont {Vos}}]{Grishina2015Nanotech}%
  \BibitemOpen
  \bibfield  {author} {\bibinfo {author} {\bibfnamefont {D.~A.}\ \bibnamefont
  {Grishina}}, \bibinfo {author} {\bibfnamefont {C.~A.~M.}\ \bibnamefont
  {Harteveld}}, \bibinfo {author} {\bibfnamefont {L.~A.}\ \bibnamefont
  {Woldering}}, \ and\ \bibinfo {author} {\bibfnamefont {W.~L.}\ \bibnamefont
  {Vos}},\ }\href {\doibase 10.1088/0957-4484/26/50/505302} {\bibfield
  {journal} {\bibinfo  {journal} {Nanotechnology}\ }\textbf {\bibinfo {volume}
  {26}},\ \bibinfo {pages} {505302} (\bibinfo {year} {2015})}\BibitemShut
  {NoStop}%
\bibitem [{\citenamefont {Vellekoop}\ and\ \citenamefont
  {Mosk}(2007)}]{Vellekoop2007OptLt}%
  \BibitemOpen
  \bibfield  {author} {\bibinfo {author} {\bibfnamefont {I.~M.}\ \bibnamefont
  {Vellekoop}}\ and\ \bibinfo {author} {\bibfnamefont {A.~P.}\ \bibnamefont
  {Mosk}},\ }\href {http://www.ncbi.nlm.nih.gov/pubmed/17700768} {\bibfield
  {journal} {\bibinfo  {journal} {Optics letters}\ }\textbf {\bibinfo {volume}
  {32}},\ \bibinfo {pages} {2309} (\bibinfo {year} {2007})}\BibitemShut
  {NoStop}%
\bibitem [{\citenamefont {Mosk}\ \emph {et~al.}(2012)\citenamefont {Mosk},
  \citenamefont {Lagendijk}, \citenamefont {Lerosey},\ and\ \citenamefont
  {Fink}}]{Mosk2012NP}%
  \BibitemOpen
  \bibfield  {author} {\bibinfo {author} {\bibfnamefont {A.~P.}\ \bibnamefont
  {Mosk}}, \bibinfo {author} {\bibfnamefont {A.}~\bibnamefont {Lagendijk}},
  \bibinfo {author} {\bibfnamefont {G.}~\bibnamefont {Lerosey}}, \ and\
  \bibinfo {author} {\bibfnamefont {M.}~\bibnamefont {Fink}},\ }\href {\doibase
  10.1038/nphoton.2012.88} {\bibfield  {journal} {\bibinfo  {journal} {Nature
  Photonics}\ }\textbf {\bibinfo {volume} {6}},\ \bibinfo {pages} {283}
  (\bibinfo {year} {2012})}\BibitemShut {NoStop}%
\bibitem [{\citenamefont {Ctistis}\ \emph {et~al.}(2010)\citenamefont
  {Ctistis}, \citenamefont {Hartsuiker}, \citenamefont {van~der Pol},
  \citenamefont {Claudon}, \citenamefont {Vos},\ and\ \citenamefont
  {G{\'{e}}rard}}]{Ctistis2010PRB}%
  \BibitemOpen
  \bibfield  {author} {\bibinfo {author} {\bibfnamefont {G.}~\bibnamefont
  {Ctistis}}, \bibinfo {author} {\bibfnamefont {A.}~\bibnamefont {Hartsuiker}},
  \bibinfo {author} {\bibfnamefont {E.}~\bibnamefont {van~der Pol}}, \bibinfo
  {author} {\bibfnamefont {J.}~\bibnamefont {Claudon}}, \bibinfo {author}
  {\bibfnamefont {W.~L.}\ \bibnamefont {Vos}}, \ and\ \bibinfo {author}
  {\bibfnamefont {J.~M.}\ \bibnamefont {G{\'{e}}rard}},\ }\href {\doibase
  10.1103/PhysRevB.82.195330} {\bibfield  {journal} {\bibinfo  {journal}
  {Physical Review B}\ }\textbf {\bibinfo {volume} {82}},\ \bibinfo {pages}
  {195330} (\bibinfo {year} {2010})}\BibitemShut {NoStop}%
\bibitem [{\citenamefont {Vos}\ \emph {et~al.}(2001)\citenamefont {Vos},
  \citenamefont {van Driel}, \citenamefont {Megens}, \citenamefont
  {Koenderink},\ and\ \citenamefont {Imhof}}]{Vos2001NATO}%
  \BibitemOpen
  \bibfield  {author} {\bibinfo {author} {\bibfnamefont {W.~L.}\ \bibnamefont
  {Vos}}, \bibinfo {author} {\bibfnamefont {H.~M.}\ \bibnamefont {van Driel}},
  \bibinfo {author} {\bibfnamefont {M.}~\bibnamefont {Megens}}, \bibinfo
  {author} {\bibfnamefont {A.~F.}\ \bibnamefont {Koenderink}}, \ and\ \bibinfo
  {author} {\bibfnamefont {A.}~\bibnamefont {Imhof}},\ }in\ \href@noop {}
  {\emph {\bibinfo {booktitle} {Photonic Crystals and Light Localization in the
  21st century}}},\ \bibinfo {editor} {edited by\ \bibinfo {editor}
  {\bibfnamefont {C.~M.}\ \bibnamefont {Soukoulis}}}\ (\bibinfo  {publisher}
  {Kluwer, Dordrecht},\ \bibinfo {year} {2001})\ pp.\ \bibinfo {pages}
  {181--198}\BibitemShut {NoStop}%
\bibitem [{\citenamefont {{Ioffe Institute (Petersburg)}}()}]{NSM}%
  \BibitemOpen
  \bibfield  {author} {\bibinfo {author} {\bibnamefont {{Ioffe Institute
  (Petersburg)}}},\ }\href@noop {} {\enquote {\bibinfo {title}
  {{Semiconductors}},}\ }\bibinfo {howpublished}
  {\url{http://www.ioffe.ru/SVA/NSM/Semicond/}}\BibitemShut {NoStop}%
\bibitem [{\citenamefont {Euser}\ \emph {et~al.}(2008)\citenamefont {Euser},
  \citenamefont {Molenaar}, \citenamefont {Fleming}, \citenamefont {Gralak},
  \citenamefont {Polman},\ and\ \citenamefont {Vos}}]{Euser2008PRB}%
  \BibitemOpen
  \bibfield  {author} {\bibinfo {author} {\bibfnamefont {T.~G.}\ \bibnamefont
  {Euser}}, \bibinfo {author} {\bibfnamefont {A.~J.}\ \bibnamefont {Molenaar}},
  \bibinfo {author} {\bibfnamefont {J.~G.}\ \bibnamefont {Fleming}}, \bibinfo
  {author} {\bibfnamefont {B.}~\bibnamefont {Gralak}}, \bibinfo {author}
  {\bibfnamefont {A.}~\bibnamefont {Polman}}, \ and\ \bibinfo {author}
  {\bibfnamefont {W.~L.}\ \bibnamefont {Vos}},\ }\href {\doibase
  10.1103/PhysRevB.77.115214} {\bibfield  {journal} {\bibinfo  {journal}
  {Physical Review B}\ }\textbf {\bibinfo {volume} {77}},\ \bibinfo {pages}
  {115214} (\bibinfo {year} {2008})}\BibitemShut {NoStop}%
\bibitem [{\citenamefont {Palacios-Lid{\'{o}}n}\ \emph
  {et~al.}(2002)\citenamefont {Palacios-Lid{\'{o}}n}, \citenamefont {Blanco},
  \citenamefont {Ibisate}, \citenamefont {Meseguer}, \citenamefont
  {L{\`{o}}pez},\ and\ \citenamefont
  {J.~S{\'{a}}nchez-Dehesa}}]{PalaciosLidon2002APL}%
  \BibitemOpen
  \bibfield  {author} {\bibinfo {author} {\bibfnamefont {E.}~\bibnamefont
  {Palacios-Lid{\'{o}}n}}, \bibinfo {author} {\bibfnamefont {A.}~\bibnamefont
  {Blanco}}, \bibinfo {author} {\bibfnamefont {M.}~\bibnamefont {Ibisate}},
  \bibinfo {author} {\bibfnamefont {F.}~\bibnamefont {Meseguer}}, \bibinfo
  {author} {\bibfnamefont {C.}~\bibnamefont {L{\`{o}}pez}}, \ and\ \bibinfo
  {author} {\bibfnamefont {J.}~\bibnamefont {J.~S{\'{a}}nchez-Dehesa}},\
  }\href@noop {} {\bibfield  {journal} {\bibinfo  {journal} {Applied Physics
  Letters}\ }\textbf {\bibinfo {volume} {81}},\ \bibinfo {pages} {4925}
  (\bibinfo {year} {2002})}\BibitemShut {NoStop}%
\bibitem [{\citenamefont {Muller}\ \emph {et~al.}(2017)\citenamefont {Muller},
  \citenamefont {Haberko}, \citenamefont {Marichy},\ and\ \citenamefont
  {Scheffold}}]{Muller2017Optica}%
  \BibitemOpen
  \bibfield  {author} {\bibinfo {author} {\bibfnamefont {N.}~\bibnamefont
  {Muller}}, \bibinfo {author} {\bibfnamefont {J.}~\bibnamefont {Haberko}},
  \bibinfo {author} {\bibfnamefont {C.}~\bibnamefont {Marichy}}, \ and\
  \bibinfo {author} {\bibfnamefont {F.}~\bibnamefont {Scheffold}},\ }\href@noop
  {} {\bibfield  {journal} {\bibinfo  {journal} {Optica}\ }\textbf {\bibinfo
  {volume} {4}},\ \bibinfo {pages} {361} (\bibinfo {year} {2017})}\BibitemShut
  {NoStop}%
\bibitem [{\citenamefont {Wijnhoven}\ \emph {et~al.}(2001)\citenamefont
  {Wijnhoven}, \citenamefont {Bechger},\ and\ \citenamefont
  {Vos}}]{Wijnhoven2001CM}%
  \BibitemOpen
  \bibfield  {author} {\bibinfo {author} {\bibfnamefont {J.~E.~G.~J.}\
  \bibnamefont {Wijnhoven}}, \bibinfo {author} {\bibfnamefont {L.}~\bibnamefont
  {Bechger}}, \ and\ \bibinfo {author} {\bibfnamefont {W.~L.}\ \bibnamefont
  {Vos}},\ }\href@noop {} {\bibfield  {journal} {\bibinfo  {journal} {Chem.
  Mater.}\ }\textbf {\bibinfo {volume} {13}},\ \bibinfo {pages} {4486}
  (\bibinfo {year} {2001})}\BibitemShut {NoStop}%
\bibitem [{\citenamefont {Wijnhoven}\ and\ \citenamefont
  {Vos}(1998)}]{Wijnhoven1998Science}%
  \BibitemOpen
  \bibfield  {author} {\bibinfo {author} {\bibfnamefont {J.~E.~G.~J.}\
  \bibnamefont {Wijnhoven}}\ and\ \bibinfo {author} {\bibfnamefont {W.~L.}\
  \bibnamefont {Vos}},\ }\href {\doibase 10.1126/science.281.5378.802}
  {\bibfield  {journal} {\bibinfo  {journal} {Science}\ }\textbf {\bibinfo
  {volume} {281}},\ \bibinfo {pages} {802} (\bibinfo {year}
  {1998})}\BibitemShut {NoStop}%
\bibitem [{\citenamefont {Furlan}\ \emph {et~al.}(2018)\citenamefont {Furlan},
  \citenamefont {Larsson}, \citenamefont {Diaz}, \citenamefont {Holler},
  \citenamefont {Krekeler}, \citenamefont {Ritter}, \citenamefont {Petrov},
  \citenamefont {Eich}, \citenamefont {Blick}, \citenamefont {Schneider},
  \citenamefont {Greving}, \citenamefont {Zierold},\ and\ \citenamefont
  {JanÃŸen}}]{Furlan2018ApplMatToday}%
  \BibitemOpen
  \bibfield  {author} {\bibinfo {author} {\bibfnamefont {K.~P.}\ \bibnamefont
  {Furlan}}, \bibinfo {author} {\bibfnamefont {E.}~\bibnamefont {Larsson}},
  \bibinfo {author} {\bibfnamefont {A.}~\bibnamefont {Diaz}}, \bibinfo {author}
  {\bibfnamefont {M.}~\bibnamefont {Holler}}, \bibinfo {author} {\bibfnamefont
  {T.}~\bibnamefont {Krekeler}}, \bibinfo {author} {\bibfnamefont
  {M.}~\bibnamefont {Ritter}}, \bibinfo {author} {\bibfnamefont {A.~Y.}\
  \bibnamefont {Petrov}}, \bibinfo {author} {\bibfnamefont {M.}~\bibnamefont
  {Eich}}, \bibinfo {author} {\bibfnamefont {R.}~\bibnamefont {Blick}},
  \bibinfo {author} {\bibfnamefont {G.~A.}\ \bibnamefont {Schneider}}, \bibinfo
  {author} {\bibfnamefont {I.}~\bibnamefont {Greving}}, \bibinfo {author}
  {\bibfnamefont {R.}~\bibnamefont {Zierold}}, \ and\ \bibinfo {author}
  {\bibfnamefont {R.}~\bibnamefont {JanÃŸen}},\ }\href {\doibase
  https://doi.org/10.1016/j.apmt.2018.10.002} {\bibfield  {journal} {\bibinfo
  {journal} {Applied Materials Today}\ }\textbf {\bibinfo {volume} {13}},\
  \bibinfo {pages} {359 } (\bibinfo {year} {2018})}\BibitemShut {NoStop}%
\bibitem [{\citenamefont {Ishizaki}\ \emph {et~al.}(2013)\citenamefont
  {Ishizaki}, \citenamefont {Koumura}, \citenamefont {Suzuki}, \citenamefont
  {Gondaira},\ and\ \citenamefont {Noda}}]{Ishizaki2013NatPhot}%
  \BibitemOpen
  \bibfield  {author} {\bibinfo {author} {\bibfnamefont {K.}~\bibnamefont
  {Ishizaki}}, \bibinfo {author} {\bibfnamefont {M.}~\bibnamefont {Koumura}},
  \bibinfo {author} {\bibfnamefont {K.}~\bibnamefont {Suzuki}}, \bibinfo
  {author} {\bibfnamefont {K.}~\bibnamefont {Gondaira}}, \ and\ \bibinfo
  {author} {\bibfnamefont {S.}~\bibnamefont {Noda}},\ }\href {\doibase
  https://doi.org/10.1038/nphoton.2012.341} {\bibfield  {journal} {\bibinfo
  {journal} {Nature Photonics}\ }\textbf {\bibinfo {volume} {7}},\ \bibinfo
  {pages} {133 } (\bibinfo {year} {2013})}\BibitemShut {NoStop}%
\bibitem [{\citenamefont {Devashish}\ \emph {et~al.}(2019)\citenamefont
  {Devashish}, \citenamefont {Ojambati}, \citenamefont {Hasan}, \citenamefont
  {van~der Vegt},\ and\ \citenamefont {Vos}}]{Devashish2019PRB}%
  \BibitemOpen
  \bibfield  {author} {\bibinfo {author} {\bibfnamefont {D.}~\bibnamefont
  {Devashish}}, \bibinfo {author} {\bibfnamefont {O.~S.}\ \bibnamefont
  {Ojambati}}, \bibinfo {author} {\bibfnamefont {S.~B.}\ \bibnamefont {Hasan}},
  \bibinfo {author} {\bibfnamefont {J.~J.~W.}\ \bibnamefont {van~der Vegt}}, \
  and\ \bibinfo {author} {\bibfnamefont {W.~L.}\ \bibnamefont {Vos}},\ }\href
  {\doibase 10.1103/PhysRevB.99.075112} {\bibfield  {journal} {\bibinfo
  {journal} {Phys. Rev. B}\ }\textbf {\bibinfo {volume} {99}},\ \bibinfo
  {pages} {075112} (\bibinfo {year} {2019})}\BibitemShut {NoStop}%
\end{thebibliography}%
\bibliographystyle{apsrev4-1}

\end{document}